\documentclass[longbibliography,aip,apl,reprint,groupedaddress,superscriptaddress]{revtex4-1} 

\usepackage{graphicx}
\usepackage{bm}
\usepackage{amsmath,amssymb}
\usepackage{hyperref}
\usepackage{lipsum}
\usepackage{color}
\definecolor{colorref}{rgb}{0.0, 0.408, 0.647}
\definecolor{grey}{rgb}{0.9, 0.9, 0.9}
\hypersetup{
	colorlinks = true,
	allcolors = colorref
}
\graphicspath{{./}}

\newcommand{\figtitle}[1]{{#1}}
\newcommand{\subfiglabel}[1]{(\MakeLowercase{#1})}
\newcommand{\rsubfiglabel}[1]{(\MakeLowercase{#1})}
\newcommand{\subpart}[1]{\textit{#1}}

\newcommand{\rfig}[1]{Fig.~\textcolor{colorref}{\ref{#1}}}

\newcommand{\req}[1]{Eq.~\textcolor{colorref}{(\ref{#1})}}
\newcommand{\rsubfig}[2]{Fig.~\textcolor{colorref}{\ref{#1}}{(\MakeLowercase{#2})}}

\newcommand{\rsubfigsA}[3]{Fig.~\textcolor{colorref}{\ref{#1}}{(\MakeLowercase{#2}),(\MakeLowercase{#3})}}

\definecolor{reviseColor}{rgb}{0, 0.0, 0.0}
\newcommand{\revise}[1]{\textcolor{reviseColor}{#1}}
\newcommand{\citeRevise}[1]{\hypersetup{citecolor=reviseColor}\cite{#1}\hypersetup{citecolor=colorref}}

\newcommand{\SeeSupply}[1]{Supplemental Material}

\usepackage[normalem]{ulem}  

\newcommand{\SIQSE}{\affiliation{1}{Shenzhen Institute for Quantum Science and Engineering, Southern University of Science and Technology, Shenzhen, Guangdong, China}}

\newcommand{\IQA}{\affiliation{3}{International Quantum Academy, Shenzhen, Guangdong, China}}
\newcommand{\GDKL}{\affiliation{4}{Guangdong Provincial Key Laboratory of Quantum Science and Engineering, Southern University of Science and Technology, Shenzhen, Guangdong, China}}
\newcommand{\HFNL}{\affiliation{5}{
Shenzhen Branch, Hefei National Laboratory, Shenzhen 518048, China}}
\newcommand{\NXU}{\affiliation{6}{
School of Physics and Electronic-Electrical Engineering, Ningxia University, Yinchuan, 750021, China}}

\begin{document}

\title{Synthetic $\pi$-flux system in 2D superconducting qubit array with tunable coupling}

\date{\today}

\author{Yiting Liu}
\thanks{These authors contributed equally to this work}
\affiliation{\SIQSE}\affiliation{\IQA}\affiliation{\GDKL}

\author{Jiawei Zhang}
\thanks{These authors contributed equally to this work}
\affiliation{\SIQSE}\affiliation{\IQA}\affiliation{\GDKL}


\author{Zechen Guo}
\affiliation{\SIQSE}\affiliation{\IQA}\affiliation{\GDKL}

\author{Peisheng Huang}
\affiliation{\NXU}\affiliation{\IQA}

\author{Wenhui Huang}
\affiliation{\SIQSE}\affiliation{\IQA}\affiliation{\GDKL}

\author{Yongqi Liang}
\affiliation{\SIQSE}\affiliation{\IQA}\affiliation{\GDKL}

\author{Jiawei Qiu}
\affiliation{\SIQSE}\affiliation{\IQA}\affiliation{\GDKL}

\author{Xuandong Sun}
\affiliation{\SIQSE}\affiliation{\IQA}\affiliation{\GDKL}

\author{Zilin Wang}
\affiliation{\NXU}\affiliation{\IQA}

\author{Changrong Xie}
\affiliation{\SIQSE}\affiliation{\IQA}\affiliation{\GDKL}

\author{Xiaohan Yang}
\affiliation{\SIQSE}\affiliation{\IQA}\affiliation{\GDKL}

\author{Jiajian Zhang}
\affiliation{\SIQSE}\affiliation{\IQA}\affiliation{\GDKL}

\author{Libo Zhang}
\affiliation{\SIQSE}\affiliation{\IQA}\affiliation{\GDKL}

\author{Ji Chu}
\affiliation{\IQA}

\author{Weijie Guo}
\affiliation{\IQA}

\author{Ji Jiang}
\affiliation{\SIQSE}\affiliation{\IQA}\affiliation{\GDKL}

\author{Xiayu Linpeng}
\affiliation{\IQA}

\author{Song Liu}
\affiliation{\SIQSE}\affiliation{\IQA}\affiliation{\GDKL}\affiliation{\HFNL}

\author{Jingjing Niu}
\affiliation{\IQA}\affiliation{\HFNL}

\author{Yuxuan Zhou}
\affiliation{\IQA}

\author{Youpeng Zhong}
\email{zhongyp@sustech.edu.cn}
\affiliation{\SIQSE}\affiliation{\IQA}\affiliation{\GDKL}\affiliation{\HFNL}

\author{Wenhui Ren}
\email{renwenhui@iqasz.cn}
\affiliation{\IQA}

\author{Ziyu Tao}
\email{taoziyu@iqasz.cn}
\affiliation{\IQA}

\author{Dapeng Yu}
\affiliation{\SIQSE}\affiliation{\IQA}\affiliation{\GDKL}\affiliation{\HFNL}

\date{\today}

\begin{abstract}
Flat-band systems provide an ideal platform for exploring exotic quantum phenomena, where the strongly suppressed kinetic energy in these flat energy bands suggests the potential for exotic phases driven by geometric structure, disorder, and interactions.
While intriguing phenomena and physical mechanisms have been unveiled in theoretical models, synthesizing such systems within scalable quantum platforms remains challenging.
Here, we present the experimental realization of a $\pi$-flux rhombic system using a two-dimensional superconducting qubit array with tunable coupling.
We experimentally observe characteristic dynamics, e.g., $\pi$-flux driven destructive interference, and demonstrate the protocol for eigenstate preparation in this rhombic array with coupler-assisted flux.
Our results %
provide future possibilities for exploring the interplay of geometry, interactions, and quantum information encoding in such degenerate systems.
\end{abstract}
\maketitle

Synthetic magnetic flux in lattice structures allows effective control of their transport properties and localization behaviors~\cite{Dalibard2011}.
For particular values of the magnetic flux, such as $\pi$-flux,
a strong localization mechanism induced by the magnetic field can be observed in two-dimensional structures, i.e., the so-called Aharonov-Bohm (AB) caging~\cite{Aharonov1959,Vidal1998,Vidal2000}.
\revise{These systems featuring flat energy bands
\citeRevise{Leykam2018,Danieli2024,Danieli2021a,Kolovsky2023}
}
are typically represented in terms of compact localized states~\cite{Bergman2008,Flach2014,Maimaiti2019,Zhang2020Ref},
in which the eigenstates are strictly confined to a specific region in real space due to destructive interference~\cite{Rhim2019}, preventing extended transport through hopping processes.
The flat bands exist in both quasi-one- and two-dimensional geometries,
including the dice~\cite{Zhang2020Ref},
Creutz~\cite{He2021}, sawtooth~\cite{Maimaiti2017},
rhombic~\cite{Creffield2010,Bermudez2011},
honeycomb~\cite{Wu2007} and Lieb lattice~\cite{Lieb1989, Julku2016}.
In particular, the rhombic lattice subjected to external magnetic fields exhibits the formation of degenerate flat energy bands with localized eigenstates confined within individual unit cells,
which have spurred significant theoretical and experimental advancements in condensed matter physics and quantum simulation
\cite{Sugawa2018,Mukherjee2018,DiLiberto2019,Kremer2020,Longhi2021,CaceresAravena2022,Li2022,Maity2024,Nicolau2023}.
Superconducting circuits have emerged as one of the promising platforms for realizing and studying lattice systems with synthetic magnetic flux
\cite{Vepsaelaeinen2019,Vepsaelaeinen2020,
Rosen2024a,Roushan2016,Wang2019,Liu2020,ZJJ2024,Martinez2023}.
Arrays of coupled transmon qubits can emulate complex tight-binding models providing unprecedented control over lattice geometry, on-site potentials, and inter-site couplings
\cite{Yan2019,Ma2019,Deng2022,Karamlou2022,Yao2023,Karamlou2024,Shi2024,Xiang2024,Deng2024,Liu2025}.
Recent experiments have demonstrated the ability to manipulate synthetic fields, enabling the realization of systems with effective magnetic flux by
parametric modulation approaches~\cite{Wang2024Science,Rosen2024a,Roushan2016,Wang2019,Liu2020,ZJJ2024}, digital gates~\cite{Neill2021},
and circuit design~\cite{Martinez2023}.
These advancements have facilitated the observation of AB caging, interaction-induced delocalization, and competition between Anderson and flat-band localization~\cite{Martinez2023,Rosen2024}, expanding our understanding of flat-band physics.
Despite these successes, the implementation of fully controlled rhombic systems incorporating synthetic fluxes within scalable superconducting quantum circuits featuring tunable couplers remains unexplored.
In this work, we experimentally realize a $\pi$-flux rhombic system using a superconducting circuit platform equipped with tunable couplers~\cite{Yan2018,Xu2020}.
By precisely engineering the synthetic magnetic flux assisted by tunable couplers, we achieve a highly controllable implementation of a flat-band system where the flux per plaquette can be tuned to $\pi$.
We further illustrate the $\pi$-flux rhombic system with anti-symmetric detunings as another effective lattice,
which enables the platforms without tunable couplings to simulate a topological trimer lattice~\cite{MartinezAlvarez2019,Anastasiadis2022} or a quasiperiodic mosaic model~\cite{Zhou2023}.
This setup enables the exploration of a synthetic $\pi$-flux system with high tunability.
Our results highlight the potential of superconducting circuits featuring tunable couplers for studying emergent phenomena and physical mechanisms in quantum systems with synthetic flux.

\begin{figure}[!t]
	\centering
    \includegraphics[width=0.46\textwidth]{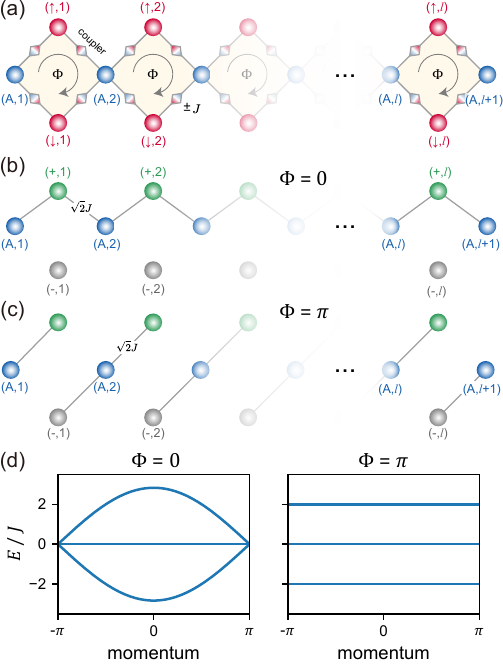}
	\caption{\label{fig1}
	\subfiglabel{a}
	Schematic of the rhombic array of superconducting qubits, where the synthetic flux $\Phi=0$ or $\Phi=\pi$ is controlled by tunable couplers (depicted as diamonds).
	\subfiglabel{b}-\subfiglabel{c}
	The effective one-dimensional model for the rhombic system: \subfiglabel{b} with $\Phi=0$, showing extended eigenstates, and \subfiglabel{c} with $\Phi=\pi$, exhibiting localized eigenstates across the lattice sites in the single-excitation subspace.
	\subfiglabel{d}
	Illustration for eigenenergy bands of the rhombic array for $\Phi=0$ (left) and $\Phi=\pi$ (right) in momentum space.
	}
\end{figure}

\subpart{Model and experimental realization.}
We consider a rhombic array of superconducting qubits with tunable nearest-neighbor (NN) couplings, as illustrated in \rsubfig{fig1}{a},
in which a synthetic flux $\Phi=0$ or $\pi$ can be effectively generated from the configuration of positive or negative coupling strengths $\pm J$ by tunable couplers.
The system can be described by the effective Hamiltonian for a rhombic lattice model with $L=3l+1$ sites:
\begin{eqnarray}
    H/\hbar = & \sum_{i,j} \Delta_{i,j} \sigma_{i,j}^{+} \sigma_{i,j}^{-}   - J \sum_{j=1}^{l} \left[
    \sigma_{A,j}^{+} \left( \sigma_{\uparrow,j}^{-} + \sigma_{\downarrow,j}^{-} \right) 
    \right. \nonumber 
    \\
     & + 
    \left. \sigma_{A,j+1}^{+} \left( \sigma_{\uparrow,j}^{-} +  e^{i\Phi} \sigma_{\downarrow,j}^{-} \right)
    + \mathrm{H.c.}\right],\label{eq:hami1}
\end{eqnarray}
where $\sigma^{+}_{i,j}$ ($\sigma_{i,j}^{-}$) represents the raising (lowering) operator for the qubit at index $(i,j)$ with $i\in \left\{A,\uparrow,\downarrow\right\}$ shown in \rsubfig{fig1}{a},
$\sigma_{i,j}^{+} \vert 0\rangle^{\otimes L} = \vert 1_{i,j}\rangle $,
$\Delta_{i,j}/2\pi$ is the frequency detuning of the qubit relative to the average system frequency,
and $J$ is the homogeneous amplitude of NN coupling strengths between qubits.
When $\Phi=0$, achieved by configuring identical positive or negative coupling strengths and setting $\Delta_{i,j}=0$,
the system in the single-excitation subspace can be mapped to an effective 1D array model with coupling strengths $\sqrt{2}J$.
In this case, the Bell state $\vert +_{j} \rangle = \left(\vert 1_{\uparrow,j} \rangle + \vert 1_{\downarrow,j}\rangle\right)/\sqrt{2}$ is regarded as a single site $(+,j)$ in the 1D array by local unitary transformations~\cite{Danieli2021}, as depicted in \rsubfig{fig1}{b}.
This effective 1D array features extended dynamics with two dispersive energy bands and
the states $\vert -_{j} \rangle = \left(\lvert 1_{\uparrow,j} \rangle - \lvert 1_{\downarrow,j} \rangle\right)/\sqrt{2}$ are zero-energy eigenstates as $H \vert -_{j} \rangle = 0$, forming a flat band at zero energy, as shown in \rsubfig{fig1}{d}.
For $\Phi=\pi$, the system in the single-excitation subspace is divided into decoupled three-level (bulk) or two-level (edge) systems depicted in \rsubfig{fig1}{c}.
This configuration gives rise to an extreme localization driven by destructive interference, characterized by oscillating populations between site $(A,j)$ and its nearest-neighboring sites~\cite{Li2022,Rosen2024}.

\begin{figure}[!t]
	\centering
    \includegraphics[width=0.48\textwidth]{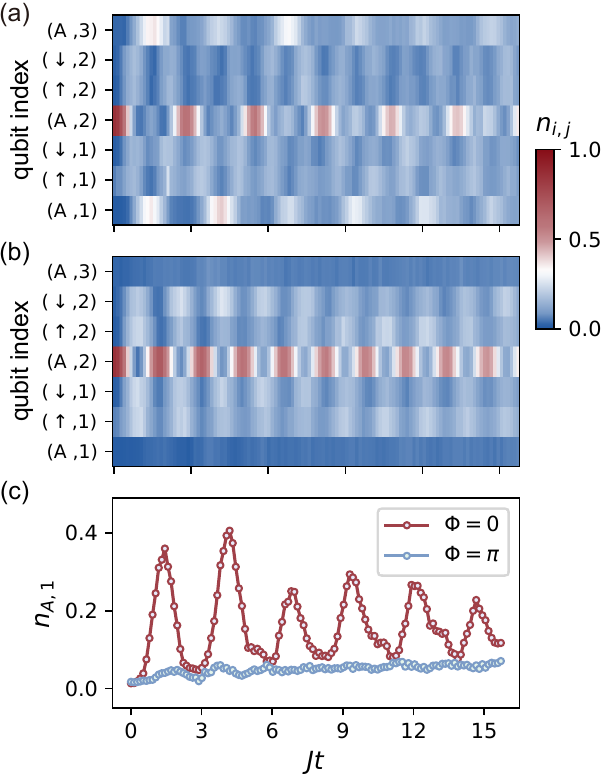}
	\caption{\label{fig2}
	\figtitle{Characteristic dynamics of the rhombic array with $\Phi=0$ and $\pi$.}
	\subfiglabel{a}-\subfiglabel{b}
	Measured population dynamics \revise{$n_{i,j}$} versus normalized evolution time $Jt$
    for \rsubfiglabel{a} $\Phi=0$ and \rsubfiglabel{b} $\Phi=\pi$.
	\subfiglabel{c}
	Time evolution of the population for qubits located at the edge sites for $\Phi=0$ and $\pi$.
	}
\end{figure}

\begin{figure*}[!t]
	\centering
    \includegraphics[width=0.99\textwidth]{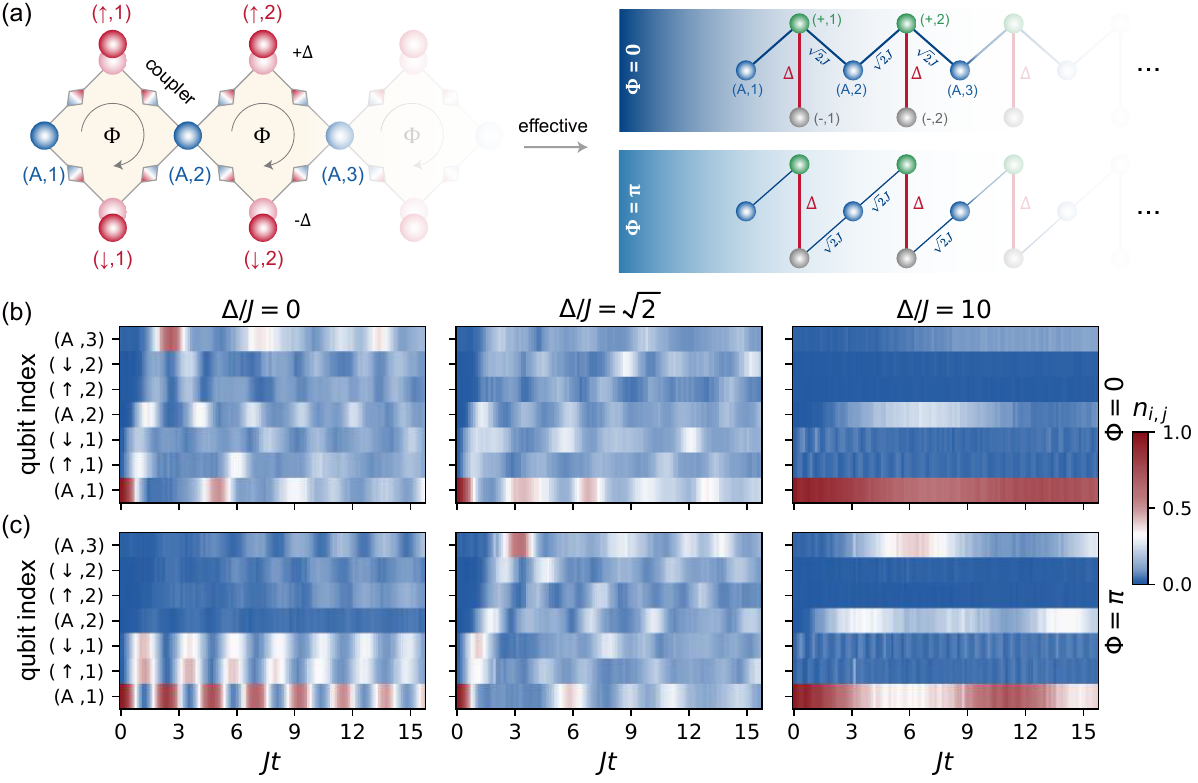}
	\caption{\label{fig3}
		\figtitle{Anti-symmetric detuning on qubit frequencies.}
		\subfiglabel{a}
		Illustration of the effective 1D model for the rhombic array with anti-symmetric detuning on qubit frequencies.
		In the basis of such a 1D system, the anti-symmetric detuning acts as an additional coupling,
        connecting the decoupled subsystems with three (two) sites in the bulk (boundary) when $\Phi=\pi$.
		\subfiglabel{b}-\subfiglabel{c}
		Measured population dynamics \revise{$n_{i,j}$} for \rsubfiglabel{b} $\Phi=0$ and \rsubfiglabel{c} $\Phi=\pi$,
		with anti-symmetric detuning $\Delta=0$, $\sqrt{2} J$ and $10 J$.
	}
\end{figure*}

Here, we demonstrate the characteristic dynamics of the rhombic array consisting of $L=7$ qubits with effective fluxes $\Phi=0$ and $\Phi=\pi$, realized using tunable couplers in superconducting quantum circuits.
As shown in \rfig{fig2}, we present the evolution of measured population dynamics $n_{i,j}(t)=\langle 1_{i,j}\rvert \rho(t) \lvert 1_{i,j}\rangle$, for individual qubits as a function of the normalized evolution time $Jt$ under the two flux configurations.
The coupling strength is set to $J\approx 2\pi \times 4.2~\mathrm{MHz}$, and the system contains $l=2$ plaquettes, corresponding to $L=7$ sites.
By initially preparing the central qubit at site $(A,2)$ in its excited state $\vert 1_{A,2}\rangle$, we examine the characteristic population dynamics of the rhombic array.
For the zero-flux $\Phi=0$ case, we configure all coupling strengths to be negative by controlling the frequencies of the tunable couplers.
The resulting population dynamics, shown in \rsubfig{fig2}{a}, exhibit oscillations traversing all sites, consistent with the picture of an effective 1D array featuring spatially extended eigenstates (as depicted in \rsubfig{fig1}{b}).
For the $\pi$-flux $\Phi=\pi$ case, we reconfigure two coupling strengths in each plaquette to be positive.
The dynamics in the $\pi$-flux rhombic system can be interpreted as a destructive interference across each $2\times 2$ plaquette,
where any initially localized excitation will be in a superposition of a finite number of localized eigenstates and remain bounded over time~\cite{Vidal1998,Martinez2023}.
In this regime, the degenerate eigenstates are localized at the site $(A,j)$ and adjacent sites in $\{(\uparrow,j)$, $(\uparrow,j-1)$, $(\downarrow,j)$, $(\downarrow,j-1)\}$.
This localization leads to vanishing population dynamics at the edge sites $(A,1)$ and $(A,3)$ when the system is initialized  in the state $\vert 1_{A,2}\rangle$,
as shown in \rsubfig{fig2}{b}.
\rsubfig{fig2}{c} further illustrates the population dynamics at the edge site $(A,1)$ as a function of the normalized evolution time $Jt$ for both configurations, $\Phi=0$ and $\pi$.

\subpart{Anti-symmetric detuning.}
We next investigate the (de)localized dynamics by introducing anti-symmetric detunings to the qubit frequencies at sites $(\uparrow,j)$ and $(\downarrow,j)$.
Without detuning, the $\pi$-flux rhombic system hosts spatially localized eigenstates at $(A,j)$ and its adjacent sites. These states form effective three- (two-) level systems on the bulk (edge) sites in the basis of $\vert \pm_{j} \rangle = \left(\vert 1_{\uparrow,j} \rangle \pm \vert 1_{\downarrow,j} \rangle\right)/\sqrt{2}$ and $\vert 1_{A,j}\rangle$ as shown in \rsubfig{fig1}{c}.
The anti-symmetric detunings on the qubit frequencies act as $\Delta\sigma^{+}_{\uparrow,j}\sigma^{-}_{\uparrow,j}$ and $-\Delta\sigma^{+}_{\downarrow,j}\sigma^{-}_{\downarrow,j}$ in the basis of $\vert 1_{\uparrow,j} \rangle$ and $\vert 1_{\downarrow,j} \rangle$, respectively,
introducing effective couplings between the states $\vert \pm_j\rangle$. This is consistent with the delocalization of the $\pi$-flux rhombus when anti-symmetric on-site disorders are introduced~\cite{Li2022,Rosen2024}.
In the $\pi$-flux rhombic array with anti-symmetric detuning $\Delta = \sqrt{2} J$, the system can be described by an effective 1D model on the sites $(A,1)$, $(+,1)$, $(-,1)$, $(A,2)$, $(+,1)\cdots$ with homogeneous couplings $\sqrt{2} J$, as illustrated in \rsubfig{fig3}{a}.
For a general value of anti-symmetric detuning $\Delta$, this $\pi$-flux system can be effectively represented by a trimer lattice \cite{MartinezAlvarez2019,Anastasiadis2022} with intra-cell couplings $\sqrt{2}J$ controlled by the tunable couplers and inter-cell couplings $\Delta$ determined by the anti-symmetric detunings on the qubit frequencies,
where the unit cell is defined on the sites $\{ (-,j-1)$, $(A,j)$, $(+,j)\}$.
According to previous theoretical studies~\cite{MartinezAlvarez2019}, the topological properties of this trimer lattice are regulated by the relative strengths of the inter-cell and intra-cell coupling amplitudes,
where the Zak phase is found to be $\mathcal{Z}=\pi$ ($\mathcal{Z}=0$) for $\Delta > \sqrt{2} J$ ($\Delta < \sqrt{2} J$) corresponding to the (non-) topological phase of the trimer lattice~\cite{MartinezAlvarez2019,Anastasiadis2022}.

We experimentally investigate a coupler-assisted rhombic system with anti-symmetric detunings on the qubit frequencies in $l=2$ plaquettes with $L=7$ sites.
A single-particle excitation is initialized at the edge site $(A,1)$, and the population dynamics of each site are tracked over normalized time $Jt$, as shown in \rsubfigsA{fig3}{b}{c}.
In the absence of anti-symmetric detuning $\Delta=0$, the zero-flux rhombic system exhibits extended population dynamics across all sites, while the $\pi$-flux rhombic system demonstrates oscillatory, localized dynamics confined to the sites $\{ (A,1), (\uparrow,1), (\downarrow,1)\}$.
When the anti-symmetric detuning is involved, the dynamics of the zero-flux system become less extended,
whereas the $\pi$-flux system exhibits increased delocalization compared to the detuning-free counterpart,
which can be effectively described using a comb-like and trimer lattice model shown in \rsubfig{fig3}{a}.
For a larger detuning $\Delta=10J$, the localized behavior in the zero-flux rhombic system is further enhanced, as presented in \rsubfig{fig3}{b}. In contrast, the $\pi$-flux system (\rsubfig{fig3}{c}) demonstrates population dynamics that traverse the middle site $(A,j)$ of the unit cell in the trimer lattice due to the sizeable inter-cell couplings~\cite{MartinezAlvarez2019,Anastasiadis2022}. Numerical simulations supporting these observations are provided in the \SeeSupply{}.

\begin{figure}[!t]
	\centering
    \includegraphics[width=0.5\textwidth]{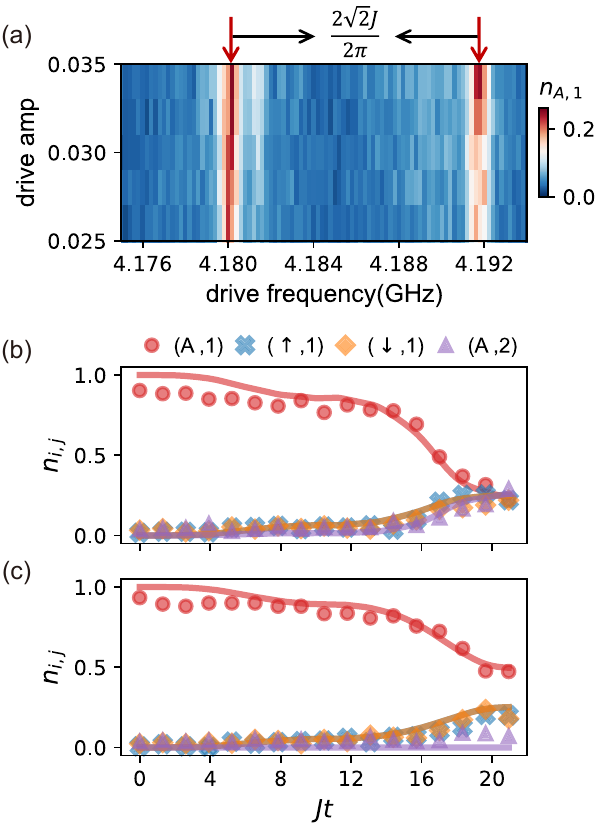}
	\caption{\label{fig4}
		\figtitle{Energy spectrum and ground state preparation of the rhombic system.}
		\subfiglabel{a}
		Spectroscopy of the $\pi$-flux rhombic array with four superconducting qubits ($Q_{A,1}$, $Q_{\uparrow,1}$, $Q_{\downarrow,1}$, $Q_{A,2}$),
		in which the increased population peaks indicate the eigenenergies of this system.
		\subfiglabel{b}-\subfiglabel{c}
		The measured population during the adiabatic preparation of the ground state for the rhombic system from the initial state $\vert 1_{A,1}\rangle$, 
        where \subfiglabel{b} $\Phi = 0$, \subfiglabel{c} $\Phi = \pi$.
        Dots indicate measured results, while solid lines denote the corresponding numerical simulation.
	}
\end{figure}

\subpart{Ground state preparation.}
We further investigate the energy spectrum and ground state of the minimal $\pi$-flux rhombic system comprising $L=4$ qubits.
To unveil the energy spectrum, we perform a spectroscopy measurement on the qubit $Q_{A,1}$ by applying a driving pulse through XY control line while keeping the qubit frequencies on resonance $\Delta_{i,j}=0$ and setting the coupling strengths according to the Hamiltonian in \req{eq:hami1}.
In this spectroscopy, an increase in the measured population indicates that the applied pump tone is in resonance with one of the eigenfrequencies of the rhombic system,
giving a transition from $\vert 0\rangle^{\otimes L}$ to the eigenstates in the single-excitation subspace~\cite{Martinez2023,Paladugu2024}.
As shown in \rsubfig{fig4}{a}, the spectroscopy reveals two population peaks
corresponding to the eigenenergies $E=\sqrt{2} J$ and $-\sqrt{2}J$ for $\pi$-flux $\Phi=\pi$.

We employ an adiabatic protocol to prepare the ground state of $\pi$-flux rhombic system~\cite{Saxberg2022,Wang2024Science}. 
The system is initialized in a simple ground state $\vert 1_{A,1} \rangle$ where the excitation is localized at the site $(A,1)$.
The frequency of this site is primarily detuned ($\vert \Delta_{A,1}\vert/J>3$) below the other qubit frequencies, and the couplings are turned off ($J=0$).
Then we gradually evolve the system by ramping up the couplings to $J\approx 2 \pi \times 4.2$ MHz and successively ramping down the detuning frequency to reach the final Hamiltonian of the rhombic system with $\Delta_{i,j}=0$.
As shown in \rsubfigsA{fig4}{b}{c}, the rhombic system follows the instantaneous ground state during the slow adiabatic evolution,
reaching the ground state with (un)equally distributed population on each site measured for $\Phi=0$ ($\Phi=\pi$),
consistent with the results calculated in numerical simulation.
\revise{
The fidelity $F = \sum_{i,j} \sqrt{n_{i,j} n_{i,j}^{\mathrm{th}}}$ for the measured population distribution $n_{i,j}$ and theoretical population distribution $n^{\mathrm{th}}_{i,j}$~\cite{Shi2023,Yan2019} gives
$F\approx 0.97$ ($F\approx 0.92$) for $\Phi=0$ ($\Phi=\pi$),
where the dominated error is mainly from the decoherence effects during the qubits evolution (see Supplemental Material for more details and numerical simulation).
}

\subpart{Discussion.}
In this work, we experimentally realize a synthetic $\pi$-flux rhombic system using tunable couplers in superconducting quantum circuits.
By properly configuring the positive or negative coupling strengths between neighboring qubits by tunable couplers,
we demonstrate the characteristic dynamics of a rhombic array of superconducting qubits with synthetic flux $\Phi=0$ and $\pi$.
We illustrate that the dynamics of $\pi$-flux rhombic system with anti-symmetric detunings can be effectively described by a trimer lattice in specific bases,
and demonstrate the system dynamics with different detunings acting as inter-cell couplings in the effective trimer lattice.
Such equivalent representation deepens our understanding of flat-band physics in the $\pi$-flux rhombic system~\cite{Danieli2021},
and provides the possibility for the experimental platforms without tunable coupling strengths to simulate a topological trimer lattice~\cite{MartinezAlvarez2019,Anastasiadis2022} or quasiperiodic mosaic model~\cite{Zhou2023} whose inter-cell couplings can be effectively controlled by the on-site detunings in real space.
Using the adiabatic protocol, we further demonstrate the ground state preparation in such a rhombic system by ramping up the tunable couplings' strengths and adjusting the qubit frequencies.
Utilizing the synthetic $\pi$-flux system with high tunability,
further efforts can be made to explore the quantum information encoding of the eigenstates in such degenerate system \cite{Roentgen2019} according to the similar protocols in recently developed dual-rail qubits
\cite{Campbell2020,Teoh2023,Chou2024,Levine2024,Koottandavida2024}.

\subpart{Acknowledgments}:
This work was supported by the Shenzhen Science and Technology Program (Grant No. RCBS20231211090824040), the National Natural Science Foundation of China (12204228, 12374474), the Innovation Program for Quantum Science and Technology (2021ZD0301703), the Shenzhen-Hong Kong Cooperation Zone for Technology and Innovation (HZQB-KCZYB-2020050), 
\revise{the Science, Technology and Innovation Commission of Shenzhen Municipality (KQTD20210811090049034)},
and Guangdong Basic and Applied Basic Research Foundation (2024A1515011714, 2022A1515110615).

\subpart{Competing interests}:
The authors declare no conflict of interest.


\begin{thebibliography}{65}%
\makeatletter
\providecommand \@ifxundefined [1]{%
 \@ifx{#1\undefined}
}%
\providecommand \@ifnum [1]{%
 \ifnum #1\expandafter \@firstoftwo
 \else \expandafter \@secondoftwo
 \fi
}%
\providecommand \@ifx [1]{%
 \ifx #1\expandafter \@firstoftwo
 \else \expandafter \@secondoftwo
 \fi
}%
\providecommand \natexlab [1]{#1}%
\providecommand \enquote  [1]{``#1''}%
\providecommand \bibnamefont  [1]{#1}%
\providecommand \bibfnamefont [1]{#1}%
\providecommand \citenamefont [1]{#1}%
\providecommand \href@noop [0]{\@secondoftwo}%
\providecommand \href [0]{\begingroup \@sanitize@url \@href}%
\providecommand \@href[1]{\@@startlink{#1}\@@href}%
\providecommand \@@href[1]{\endgroup#1\@@endlink}%
\providecommand \@sanitize@url [0]{\catcode `\\12\catcode `\$12\catcode
  `\&12\catcode `\#12\catcode `\^12\catcode `\_12\catcode `\%12\relax}%
\providecommand \@@startlink[1]{}%
\providecommand \@@endlink[0]{}%
\providecommand \url  [0]{\begingroup\@sanitize@url \@url }%
\providecommand \@url [1]{\endgroup\@href {#1}{\urlprefix }}%
\providecommand \urlprefix  [0]{URL }%
\providecommand \Eprint [0]{\href }%
\providecommand \doibase [0]{https://doi.org/}%
\providecommand \selectlanguage [0]{\@gobble}%
\providecommand \bibinfo  [0]{\@secondoftwo}%
\providecommand \bibfield  [0]{\@secondoftwo}%
\providecommand \translation [1]{[#1]}%
\providecommand \BibitemOpen [0]{}%
\providecommand \bibitemStop [0]{}%
\providecommand \bibitemNoStop [0]{.\EOS\space}%
\providecommand \EOS [0]{\spacefactor3000\relax}%
\providecommand \BibitemShut  [1]{\csname bibitem#1\endcsname}%
\let\auto@bib@innerbib\@empty
\bibitem [{\citenamefont {Dalibard}\ \emph {et~al.}(2011)\citenamefont
  {Dalibard}, \citenamefont {Gerbier}, \citenamefont {Juzeliūnas},\ and\
  \citenamefont {Öhberg}}]{Dalibard2011}%
  \BibitemOpen
  \bibfield  {author} {\bibinfo {author} {\bibfnamefont {J.}~\bibnamefont
  {Dalibard}}, \bibinfo {author} {\bibfnamefont {F.}~\bibnamefont {Gerbier}},
  \bibinfo {author} {\bibfnamefont {G.}~\bibnamefont {Juzeliūnas}},\ and\
  \bibinfo {author} {\bibfnamefont {P.}~\bibnamefont {Öhberg}},\ }\bibfield
  {title} {\emph {\bibinfo {title} {Colloquium: Artificial gauge potentials for
  neutral atoms}},\ }\href {https://doi.org/10.1103/revmodphys.83.1523}
  {\bibfield  {journal} {\bibinfo  {journal} {Rev. Mod. Phys.}\ }\textbf
  {\bibinfo {volume} {83}},\ \bibinfo {pages} {1523} (\bibinfo {year}
  {2011})}\BibitemShut {NoStop}%
\bibitem [{\citenamefont {Aharonov}\ and\ \citenamefont
  {Bohm}(1959)}]{Aharonov1959}%
  \BibitemOpen
  \bibfield  {author} {\bibinfo {author} {\bibfnamefont {Y.}~\bibnamefont
  {Aharonov}}\ and\ \bibinfo {author} {\bibfnamefont {D.}~\bibnamefont
  {Bohm}},\ }\bibfield  {title} {\emph {\bibinfo {title} {Significance of
  electromagnetic potentials in the quantum theory}},\ }\href
  {https://doi.org/10.1103/physrev.115.485} {\bibfield  {journal} {\bibinfo
  {journal} {Phys. Rev.}\ }\textbf {\bibinfo {volume} {115}},\ \bibinfo {pages}
  {485} (\bibinfo {year} {1959})}\BibitemShut {NoStop}%
\bibitem [{\citenamefont {Vidal}\ \emph {et~al.}(1998)\citenamefont {Vidal},
  \citenamefont {Mosseri},\ and\ \citenamefont {Douçot}}]{Vidal1998}%
  \BibitemOpen
  \bibfield  {author} {\bibinfo {author} {\bibfnamefont {J.}~\bibnamefont
  {Vidal}}, \bibinfo {author} {\bibfnamefont {R.}~\bibnamefont {Mosseri}},\
  and\ \bibinfo {author} {\bibfnamefont {B.}~\bibnamefont {Douçot}},\
  }\bibfield  {title} {\emph {\bibinfo {title} {Aharonov-{Bohm} cages in
  two-dimensional structures}},\ }\href
  {https://doi.org/10.1103/physrevlett.81.5888} {\bibfield  {journal} {\bibinfo
   {journal} {Phys. Rev. Lett.}\ }\textbf {\bibinfo {volume} {81}},\ \bibinfo
  {pages} {5888} (\bibinfo {year} {1998})}\BibitemShut {NoStop}%
\bibitem [{\citenamefont {Vidal}\ \emph {et~al.}(2000)\citenamefont {Vidal},
  \citenamefont {Douçot}, \citenamefont {Mosseri},\ and\ \citenamefont
  {Butaud}}]{Vidal2000}%
  \BibitemOpen
  \bibfield  {author} {\bibinfo {author} {\bibfnamefont {J.}~\bibnamefont
  {Vidal}}, \bibinfo {author} {\bibfnamefont {B.}~\bibnamefont {Douçot}},
  \bibinfo {author} {\bibfnamefont {R.}~\bibnamefont {Mosseri}},\ and\ \bibinfo
  {author} {\bibfnamefont {P.}~\bibnamefont {Butaud}},\ }\bibfield  {title}
  {\emph {\bibinfo {title} {Interaction induced delocalization for two
  particles in a periodic potential}},\ }\href
  {https://doi.org/10.1103/physrevlett.85.3906} {\bibfield  {journal} {\bibinfo
   {journal} {Phys. Rev. Lett.}\ }\textbf {\bibinfo {volume} {85}},\ \bibinfo
  {pages} {3906} (\bibinfo {year} {2000})}\BibitemShut {NoStop}%
\bibitem [{\citenamefont {Leykam}\ \emph {et~al.}(2018)\citenamefont {Leykam},
  \citenamefont {Andreanov},\ and\ \citenamefont {Flach}}]{Leykam2018}%
  \BibitemOpen
  \bibfield  {author} {\bibinfo {author} {\bibfnamefont {D.}~\bibnamefont
  {Leykam}}, \bibinfo {author} {\bibfnamefont {A.}~\bibnamefont {Andreanov}},\
  and\ \bibinfo {author} {\bibfnamefont {S.}~\bibnamefont {Flach}},\ }\bibfield
   {title} {\emph {\bibinfo {title} {\color{reviseColor}{Artificial flat band
  systems: from lattice models to experiments}}},\ }\href
  {https://doi.org/10.1080/23746149.2018.1473052} {\bibfield  {journal}
  {\bibinfo  {journal} {Advances in Physics: X}\ }\textbf {\bibinfo {volume}
  {3}},\ \bibinfo {pages} {1473052} (\bibinfo {year} {2018})}\BibitemShut
  {NoStop}%
\bibitem [{\citenamefont {Danieli}\ \emph {et~al.}(2024)\citenamefont
  {Danieli}, \citenamefont {Andreanov}, \citenamefont {Leykam},\ and\
  \citenamefont {Flach}}]{Danieli2024}%
  \BibitemOpen
  \bibfield  {author} {\bibinfo {author} {\bibfnamefont {C.}~\bibnamefont
  {Danieli}}, \bibinfo {author} {\bibfnamefont {A.}~\bibnamefont {Andreanov}},
  \bibinfo {author} {\bibfnamefont {D.}~\bibnamefont {Leykam}},\ and\ \bibinfo
  {author} {\bibfnamefont {S.}~\bibnamefont {Flach}},\ }\bibfield  {title}
  {\emph {\bibinfo {title} {\color{reviseColor}{Flat band fine-tuning and its
  photonic applications}}},\ }\href {https://doi.org/10.1515/nanoph-2024-0135}
  {\bibfield  {journal} {\bibinfo  {journal} {Nanophotonics}\ }\textbf
  {\bibinfo {volume} {13}},\ \bibinfo {pages} {3925} (\bibinfo {year}
  {2024})}\BibitemShut {NoStop}%
\bibitem [{\citenamefont {Danieli}\ \emph
  {et~al.}(2021{\natexlab{a}})\citenamefont {Danieli}, \citenamefont
  {Andreanov}, \citenamefont {Mithun},\ and\ \citenamefont
  {Flach}}]{Danieli2021a}%
  \BibitemOpen
  \bibfield  {author} {\bibinfo {author} {\bibfnamefont {C.}~\bibnamefont
  {Danieli}}, \bibinfo {author} {\bibfnamefont {A.}~\bibnamefont {Andreanov}},
  \bibinfo {author} {\bibfnamefont {T.}~\bibnamefont {Mithun}},\ and\ \bibinfo
  {author} {\bibfnamefont {S.}~\bibnamefont {Flach}},\ }\bibfield  {title}
  {\emph {\bibinfo {title} {\color{reviseColor}{{Quantum} caging in interacting
  many-body all-bands-flat lattices}}},\ }\href
  {https://doi.org/10.1103/physrevb.104.085132} {\bibfield  {journal} {\bibinfo
   {journal} {Phys. Rev. B}\ }\textbf {\bibinfo {volume} {104}},\ \bibinfo
  {pages} {085132} (\bibinfo {year} {2021}{\natexlab{a}})}\BibitemShut
  {NoStop}%
\bibitem [{\citenamefont {Kolovsky}\ \emph {et~al.}(2023)\citenamefont
  {Kolovsky}, \citenamefont {Muraev},\ and\ \citenamefont
  {Flach}}]{Kolovsky2023}%
  \BibitemOpen
  \bibfield  {author} {\bibinfo {author} {\bibfnamefont {A.~R.}\ \bibnamefont
  {Kolovsky}}, \bibinfo {author} {\bibfnamefont {P.~S.}\ \bibnamefont
  {Muraev}},\ and\ \bibinfo {author} {\bibfnamefont {S.}~\bibnamefont
  {Flach}},\ }\bibfield  {title} {\emph {\bibinfo {title}
  {\color{reviseColor}{Conductance transition with interacting bosons in an
  Aharonov-Bohm cage}}},\ }\href {https://doi.org/10.1103/physreva.108.l010201}
  {\bibfield  {journal} {\bibinfo  {journal} {Phys. Rev. A}\ }\textbf {\bibinfo
  {volume} {108}},\ \bibinfo {pages} {L010201} (\bibinfo {year}
  {2023})}\BibitemShut {NoStop}%
\bibitem [{\citenamefont {Bergman}\ \emph {et~al.}(2008)\citenamefont
  {Bergman}, \citenamefont {Wu},\ and\ \citenamefont {Balents}}]{Bergman2008}%
  \BibitemOpen
  \bibfield  {author} {\bibinfo {author} {\bibfnamefont {D.~L.}\ \bibnamefont
  {Bergman}}, \bibinfo {author} {\bibfnamefont {C.}~\bibnamefont {Wu}},\ and\
  \bibinfo {author} {\bibfnamefont {L.}~\bibnamefont {Balents}},\ }\bibfield
  {title} {\emph {\bibinfo {title} {Band touching from real-space topology in
  frustrated hopping models}},\ }\href
  {https://doi.org/10.1103/physrevb.78.125104} {\bibfield  {journal} {\bibinfo
  {journal} {Phys. Rev. B}\ }\textbf {\bibinfo {volume} {78}},\ \bibinfo
  {pages} {125104} (\bibinfo {year} {2008})}\BibitemShut {NoStop}%
\bibitem [{\citenamefont {Flach}\ \emph {et~al.}(2014)\citenamefont {Flach},
  \citenamefont {Leykam}, \citenamefont {Bodyfelt}, \citenamefont {Matthies},\
  and\ \citenamefont {Desyatnikov}}]{Flach2014}%
  \BibitemOpen
  \bibfield  {author} {\bibinfo {author} {\bibfnamefont {S.}~\bibnamefont
  {Flach}}, \bibinfo {author} {\bibfnamefont {D.}~\bibnamefont {Leykam}},
  \bibinfo {author} {\bibfnamefont {J.~D.}\ \bibnamefont {Bodyfelt}}, \bibinfo
  {author} {\bibfnamefont {P.}~\bibnamefont {Matthies}},\ and\ \bibinfo
  {author} {\bibfnamefont {A.~S.}\ \bibnamefont {Desyatnikov}},\ }\bibfield
  {title} {\emph {\bibinfo {title} {Detangling flat bands into {Fano}
  lattices}},\ }\href {https://doi.org/10.1209/0295-5075/105/30001} {\bibfield
  {journal} {\bibinfo  {journal} {Europhys. Lett.}\ }\textbf {\bibinfo {volume}
  {105}},\ \bibinfo {pages} {30001} (\bibinfo {year} {2014})}\BibitemShut
  {NoStop}%
\bibitem [{\citenamefont {Maimaiti}\ \emph {et~al.}(2019)\citenamefont
  {Maimaiti}, \citenamefont {Flach},\ and\ \citenamefont
  {Andreanov}}]{Maimaiti2019}%
  \BibitemOpen
  \bibfield  {author} {\bibinfo {author} {\bibfnamefont {W.}~\bibnamefont
  {Maimaiti}}, \bibinfo {author} {\bibfnamefont {S.}~\bibnamefont {Flach}},\
  and\ \bibinfo {author} {\bibfnamefont {A.}~\bibnamefont {Andreanov}},\
  }\bibfield  {title} {\emph {\bibinfo {title} {Universal $d=1$ flat band
  generator from compact localized states}},\ }\href
  {https://doi.org/10.1103/physrevb.99.125129} {\bibfield  {journal} {\bibinfo
  {journal} {Phys. Rev. B}\ }\textbf {\bibinfo {volume} {99}},\ \bibinfo
  {pages} {125129} (\bibinfo {year} {2019})}\BibitemShut {NoStop}%
\bibitem [{\citenamefont {Zhang}\ and\ \citenamefont
  {Jin}(2020)}]{Zhang2020Ref}%
  \BibitemOpen
  \bibfield  {author} {\bibinfo {author} {\bibfnamefont {S.~M.}\ \bibnamefont
  {Zhang}}\ and\ \bibinfo {author} {\bibfnamefont {L.}~\bibnamefont {Jin}},\
  }\bibfield  {title} {\emph {\bibinfo {title} {Compact localized states and
  localization dynamics in the dice lattice}},\ }\href
  {https://doi.org/10.1103/physrevb.102.054301} {\bibfield  {journal} {\bibinfo
   {journal} {Phys. Rev. B}\ }\textbf {\bibinfo {volume} {102}},\ \bibinfo
  {pages} {054301} (\bibinfo {year} {2020})}\BibitemShut {NoStop}%
\bibitem [{\citenamefont {Rhim}\ and\ \citenamefont {Yang}(2019)}]{Rhim2019}%
  \BibitemOpen
  \bibfield  {author} {\bibinfo {author} {\bibfnamefont {J.-W.}\ \bibnamefont
  {Rhim}}\ and\ \bibinfo {author} {\bibfnamefont {B.-J.}\ \bibnamefont
  {Yang}},\ }\bibfield  {title} {\emph {\bibinfo {title} {Classification of
  flat bands according to the band-crossing singularity of {Bloch} wave
  functions}},\ }\href {https://doi.org/10.1103/physrevb.99.045107} {\bibfield
  {journal} {\bibinfo  {journal} {Phys. Rev. B}\ }\textbf {\bibinfo {volume}
  {99}},\ \bibinfo {pages} {045107} (\bibinfo {year} {2019})}\BibitemShut
  {NoStop}%
\bibitem [{\citenamefont {He}\ \emph {et~al.}(2021)\citenamefont {He},
  \citenamefont {Mao}, \citenamefont {Cai}, \citenamefont {Zhang},
  \citenamefont {Li}, \citenamefont {Yuan}, \citenamefont {Zhu},\ and\
  \citenamefont {Wang}}]{He2021}%
  \BibitemOpen
  \bibfield  {author} {\bibinfo {author} {\bibfnamefont {Y.}~\bibnamefont
  {He}}, \bibinfo {author} {\bibfnamefont {R.}~\bibnamefont {Mao}}, \bibinfo
  {author} {\bibfnamefont {H.}~\bibnamefont {Cai}}, \bibinfo {author}
  {\bibfnamefont {J.-X.}\ \bibnamefont {Zhang}}, \bibinfo {author}
  {\bibfnamefont {Y.}~\bibnamefont {Li}}, \bibinfo {author} {\bibfnamefont
  {L.}~\bibnamefont {Yuan}}, \bibinfo {author} {\bibfnamefont {S.-Y.}\
  \bibnamefont {Zhu}},\ and\ \bibinfo {author} {\bibfnamefont {D.-W.}\
  \bibnamefont {Wang}},\ }\bibfield  {title} {\emph {\bibinfo {title}
  {Flat-band localization in {Creutz} superradiance lattices}},\ }\href
  {https://doi.org/10.1103/physrevlett.126.103601} {\bibfield  {journal}
  {\bibinfo  {journal} {Phys. Rev. Lett.}\ }\textbf {\bibinfo {volume} {126}},\
  \bibinfo {pages} {103601} (\bibinfo {year} {2021})}\BibitemShut {NoStop}%
\bibitem [{\citenamefont {Maimaiti}\ \emph {et~al.}(2017)\citenamefont
  {Maimaiti}, \citenamefont {Andreanov}, \citenamefont {Park}, \citenamefont
  {Gendelman},\ and\ \citenamefont {Flach}}]{Maimaiti2017}%
  \BibitemOpen
  \bibfield  {author} {\bibinfo {author} {\bibfnamefont {W.}~\bibnamefont
  {Maimaiti}}, \bibinfo {author} {\bibfnamefont {A.}~\bibnamefont {Andreanov}},
  \bibinfo {author} {\bibfnamefont {H.~C.}\ \bibnamefont {Park}}, \bibinfo
  {author} {\bibfnamefont {O.}~\bibnamefont {Gendelman}},\ and\ \bibinfo
  {author} {\bibfnamefont {S.}~\bibnamefont {Flach}},\ }\bibfield  {title}
  {\emph {\bibinfo {title} {Compact localized states and flat-band generators
  in one dimension}},\ }\href {https://doi.org/10.1103/physrevb.95.115135}
  {\bibfield  {journal} {\bibinfo  {journal} {Phys. Rev. B}\ }\textbf {\bibinfo
  {volume} {95}},\ \bibinfo {pages} {115135} (\bibinfo {year}
  {2017})}\BibitemShut {NoStop}%
\bibitem [{\citenamefont {Creffield}\ and\ \citenamefont
  {Platero}(2010)}]{Creffield2010}%
  \BibitemOpen
  \bibfield  {author} {\bibinfo {author} {\bibfnamefont {C.~E.}\ \bibnamefont
  {Creffield}}\ and\ \bibinfo {author} {\bibfnamefont {G.}~\bibnamefont
  {Platero}},\ }\bibfield  {title} {\emph {\bibinfo {title} {Coherent control
  of interacting particles using dynamical and {Aharonov}-{Bohm} phases}},\
  }\href {https://doi.org/10.1103/physrevlett.105.086804} {\bibfield  {journal}
  {\bibinfo  {journal} {Phys. Rev. Lett.}\ }\textbf {\bibinfo {volume} {105}},\
  \bibinfo {pages} {086804} (\bibinfo {year} {2010})}\BibitemShut {NoStop}%
\bibitem [{\citenamefont {Bermudez}\ \emph {et~al.}(2011)\citenamefont
  {Bermudez}, \citenamefont {Schaetz},\ and\ \citenamefont
  {Porras}}]{Bermudez2011}%
  \BibitemOpen
  \bibfield  {author} {\bibinfo {author} {\bibfnamefont {A.}~\bibnamefont
  {Bermudez}}, \bibinfo {author} {\bibfnamefont {T.}~\bibnamefont {Schaetz}},\
  and\ \bibinfo {author} {\bibfnamefont {D.}~\bibnamefont {Porras}},\
  }\bibfield  {title} {\emph {\bibinfo {title} {Synthetic gauge fields for
  vibrational excitations of trapped ions}},\ }\href
  {https://doi.org/10.1103/physrevlett.107.150501} {\bibfield  {journal}
  {\bibinfo  {journal} {Phys. Rev. Lett.}\ }\textbf {\bibinfo {volume} {107}},\
  \bibinfo {pages} {150501} (\bibinfo {year} {2011})}\BibitemShut {NoStop}%
\bibitem [{\citenamefont {Wu}\ \emph {et~al.}(2007)\citenamefont {Wu},
  \citenamefont {Bergman}, \citenamefont {Balents},\ and\ \citenamefont
  {Das~Sarma}}]{Wu2007}%
  \BibitemOpen
  \bibfield  {author} {\bibinfo {author} {\bibfnamefont {C.}~\bibnamefont
  {Wu}}, \bibinfo {author} {\bibfnamefont {D.}~\bibnamefont {Bergman}},
  \bibinfo {author} {\bibfnamefont {L.}~\bibnamefont {Balents}},\ and\ \bibinfo
  {author} {\bibfnamefont {S.}~\bibnamefont {Das~Sarma}},\ }\bibfield  {title}
  {\emph {\bibinfo {title} {Flat bands and {Wigner} crystallization in the
  honeycomb optical lattice}},\ }\href
  {https://doi.org/10.1103/physrevlett.99.070401} {\bibfield  {journal}
  {\bibinfo  {journal} {Phys. Rev. Lett.}\ }\textbf {\bibinfo {volume} {99}},\
  \bibinfo {pages} {070401} (\bibinfo {year} {2007})}\BibitemShut {NoStop}%
\bibitem [{\citenamefont {Lieb}(1989)}]{Lieb1989}%
  \BibitemOpen
  \bibfield  {author} {\bibinfo {author} {\bibfnamefont {E.~H.}\ \bibnamefont
  {Lieb}},\ }\bibfield  {title} {\emph {\bibinfo {title} {Two theorems on the
  {Hubbard} model}},\ }\href {https://doi.org/10.1103/physrevlett.62.1201}
  {\bibfield  {journal} {\bibinfo  {journal} {Phys. Rev. Lett.}\ }\textbf
  {\bibinfo {volume} {62}},\ \bibinfo {pages} {1201} (\bibinfo {year}
  {1989})}\BibitemShut {NoStop}%
\bibitem [{\citenamefont {Julku}\ \emph {et~al.}(2016)\citenamefont {Julku},
  \citenamefont {Peotta}, \citenamefont {Vanhala}, \citenamefont {Kim},\ and\
  \citenamefont {Törmä}}]{Julku2016}%
  \BibitemOpen
  \bibfield  {author} {\bibinfo {author} {\bibfnamefont {A.}~\bibnamefont
  {Julku}}, \bibinfo {author} {\bibfnamefont {S.}~\bibnamefont {Peotta}},
  \bibinfo {author} {\bibfnamefont {T.~I.}\ \bibnamefont {Vanhala}}, \bibinfo
  {author} {\bibfnamefont {D.-H.}\ \bibnamefont {Kim}},\ and\ \bibinfo {author}
  {\bibfnamefont {P.}~\bibnamefont {Törmä}},\ }\bibfield  {title} {\emph
  {\bibinfo {title} {Geometric origin of superfluidity in the {Lieb}-lattice
  flat band}},\ }\href {https://doi.org/10.1103/physrevlett.117.045303}
  {\bibfield  {journal} {\bibinfo  {journal} {Phys. Rev. Lett.}\ }\textbf
  {\bibinfo {volume} {117}},\ \bibinfo {pages} {045303} (\bibinfo {year}
  {2016})}\BibitemShut {NoStop}%
\bibitem [{\citenamefont {Sugawa}\ \emph {et~al.}(2018)\citenamefont {Sugawa},
  \citenamefont {Salces-Carcoba}, \citenamefont {Perry}, \citenamefont {Yue},\
  and\ \citenamefont {Spielman}}]{Sugawa2018}%
  \BibitemOpen
  \bibfield  {author} {\bibinfo {author} {\bibfnamefont {S.}~\bibnamefont
  {Sugawa}}, \bibinfo {author} {\bibfnamefont {F.}~\bibnamefont
  {Salces-Carcoba}}, \bibinfo {author} {\bibfnamefont {A.~R.}\ \bibnamefont
  {Perry}}, \bibinfo {author} {\bibfnamefont {Y.}~\bibnamefont {Yue}},\ and\
  \bibinfo {author} {\bibfnamefont {I.~B.}\ \bibnamefont {Spielman}},\
  }\bibfield  {title} {\emph {\bibinfo {title} {Second {Chern} number of a
  quantum-simulated non-{Abelian} {Yang} monopole}},\ }\href
  {https://doi.org/10.1126/science.aam9031} {\bibfield  {journal} {\bibinfo
  {journal} {Science}\ }\textbf {\bibinfo {volume} {360}},\ \bibinfo {pages}
  {1429} (\bibinfo {year} {2018})}\BibitemShut {NoStop}%
\bibitem [{\citenamefont {Mukherjee}\ \emph {et~al.}(2018)\citenamefont
  {Mukherjee}, \citenamefont {Di~Liberto}, \citenamefont {Öhberg},
  \citenamefont {Thomson},\ and\ \citenamefont {Goldman}}]{Mukherjee2018}%
  \BibitemOpen
  \bibfield  {author} {\bibinfo {author} {\bibfnamefont {S.}~\bibnamefont
  {Mukherjee}}, \bibinfo {author} {\bibfnamefont {M.}~\bibnamefont
  {Di~Liberto}}, \bibinfo {author} {\bibfnamefont {P.}~\bibnamefont {Öhberg}},
  \bibinfo {author} {\bibfnamefont {R.~R.}\ \bibnamefont {Thomson}},\ and\
  \bibinfo {author} {\bibfnamefont {N.}~\bibnamefont {Goldman}},\ }\bibfield
  {title} {\emph {\bibinfo {title} {Experimental observation of
  {Aharonov}-{Bohm} cages in photonic lattices}},\ }\href
  {https://doi.org/10.1103/physrevlett.121.075502} {\bibfield  {journal}
  {\bibinfo  {journal} {Phys. Rev. Lett.}\ }\textbf {\bibinfo {volume} {121}},\
  \bibinfo {pages} {075502} (\bibinfo {year} {2018})}\BibitemShut {NoStop}%
\bibitem [{\citenamefont {Di~Liberto}\ \emph {et~al.}(2019)\citenamefont
  {Di~Liberto}, \citenamefont {Mukherjee},\ and\ \citenamefont
  {Goldman}}]{DiLiberto2019}%
  \BibitemOpen
  \bibfield  {author} {\bibinfo {author} {\bibfnamefont {M.}~\bibnamefont
  {Di~Liberto}}, \bibinfo {author} {\bibfnamefont {S.}~\bibnamefont
  {Mukherjee}},\ and\ \bibinfo {author} {\bibfnamefont {N.}~\bibnamefont
  {Goldman}},\ }\bibfield  {title} {\emph {\bibinfo {title} {Nonlinear dynamics
  of {Aharonov}-{Bohm} cages}},\ }\href
  {https://doi.org/10.1103/PhysRevA.100.043829} {\bibfield  {journal} {\bibinfo
   {journal} {Phys. Rev. A}\ }\textbf {\bibinfo {volume} {100}},\ \bibinfo
  {pages} {043829} (\bibinfo {year} {2019})}\BibitemShut {NoStop}%
\bibitem [{\citenamefont {Kremer}\ \emph {et~al.}(2020)\citenamefont {Kremer},
  \citenamefont {Petrides}, \citenamefont {Meyer}, \citenamefont {Heinrich},
  \citenamefont {Zilberberg},\ and\ \citenamefont {Szameit}}]{Kremer2020}%
  \BibitemOpen
  \bibfield  {author} {\bibinfo {author} {\bibfnamefont {M.}~\bibnamefont
  {Kremer}}, \bibinfo {author} {\bibfnamefont {I.}~\bibnamefont {Petrides}},
  \bibinfo {author} {\bibfnamefont {E.}~\bibnamefont {Meyer}}, \bibinfo
  {author} {\bibfnamefont {M.}~\bibnamefont {Heinrich}}, \bibinfo {author}
  {\bibfnamefont {O.}~\bibnamefont {Zilberberg}},\ and\ \bibinfo {author}
  {\bibfnamefont {A.}~\bibnamefont {Szameit}},\ }\bibfield  {title} {\emph
  {\bibinfo {title} {A square-root topological insulator with non-quantized
  indices realized with photonic {Aharonov}-{Bohm} cages}},\ }\href
  {https://doi.org/10.1038/s41467-020-14692-4} {\bibfield  {journal} {\bibinfo
  {journal} {Nat. Commun.}\ }\textbf {\bibinfo {volume} {11}},\ \bibinfo
  {pages} {907} (\bibinfo {year} {2020})}\BibitemShut {NoStop}%
\bibitem [{\citenamefont {Longhi}(2021)}]{Longhi2021}%
  \BibitemOpen
  \bibfield  {author} {\bibinfo {author} {\bibfnamefont {S.}~\bibnamefont
  {Longhi}},\ }\bibfield  {title} {\emph {\bibinfo {title} {Inverse {Anderson}
  transition in photonic cages}},\ }\href {https://doi.org/10.1364/ol.430196}
  {\bibfield  {journal} {\bibinfo  {journal} {Opt. Lett.}\ }\textbf {\bibinfo
  {volume} {46}},\ \bibinfo {pages} {2872} (\bibinfo {year}
  {2021})}\BibitemShut {NoStop}%
\bibitem [{\citenamefont {Cáceres-Aravena}\ \emph {et~al.}(2022)\citenamefont
  {Cáceres-Aravena}, \citenamefont {Guzmán-Silva}, \citenamefont {Salinas},\
  and\ \citenamefont {Vicencio}}]{CaceresAravena2022}%
  \BibitemOpen
  \bibfield  {author} {\bibinfo {author} {\bibfnamefont {G.}~\bibnamefont
  {Cáceres-Aravena}}, \bibinfo {author} {\bibfnamefont {D.}~\bibnamefont
  {Guzmán-Silva}}, \bibinfo {author} {\bibfnamefont {I.}~\bibnamefont
  {Salinas}},\ and\ \bibinfo {author} {\bibfnamefont {R.~A.}\ \bibnamefont
  {Vicencio}},\ }\bibfield  {title} {\emph {\bibinfo {title} {Controlled
  transport based on multiorbital {Aharonov}-{Bohm} photonic caging}},\ }\href
  {https://doi.org/10.1103/physrevlett.128.256602} {\bibfield  {journal}
  {\bibinfo  {journal} {Phys. Rev. Lett.}\ }\textbf {\bibinfo {volume} {128}},\
  \bibinfo {pages} {256602} (\bibinfo {year} {2022})}\BibitemShut {NoStop}%
\bibitem [{\citenamefont {Li}\ \emph {et~al.}(2022)\citenamefont {Li},
  \citenamefont {Dong}, \citenamefont {Longhi}, \citenamefont {Liang},
  \citenamefont {Xie},\ and\ \citenamefont {Yan}}]{Li2022}%
  \BibitemOpen
  \bibfield  {author} {\bibinfo {author} {\bibfnamefont {H.}~\bibnamefont
  {Li}}, \bibinfo {author} {\bibfnamefont {Z.}~\bibnamefont {Dong}}, \bibinfo
  {author} {\bibfnamefont {S.}~\bibnamefont {Longhi}}, \bibinfo {author}
  {\bibfnamefont {Q.}~\bibnamefont {Liang}}, \bibinfo {author} {\bibfnamefont
  {D.}~\bibnamefont {Xie}},\ and\ \bibinfo {author} {\bibfnamefont
  {B.}~\bibnamefont {Yan}},\ }\bibfield  {title} {\emph {\bibinfo {title}
  {Aharonov-{Bohm} caging and inverse {Anderson} transition in ultracold
  atoms}},\ }\href {https://doi.org/10.1103/physrevlett.129.220403} {\bibfield
  {journal} {\bibinfo  {journal} {Phys. Rev. Lett.}\ }\textbf {\bibinfo
  {volume} {129}},\ \bibinfo {pages} {220403} (\bibinfo {year}
  {2022})}\BibitemShut {NoStop}%
\bibitem [{\citenamefont {Maity}\ \emph {et~al.}(2024)\citenamefont {Maity},
  \citenamefont {Paul}, \citenamefont {Sharma},\ and\ \citenamefont
  {Mishra}}]{Maity2024}%
  \BibitemOpen
  \bibfield  {author} {\bibinfo {author} {\bibfnamefont {S.}~\bibnamefont
  {Maity}}, \bibinfo {author} {\bibfnamefont {B.}~\bibnamefont {Paul}},
  \bibinfo {author} {\bibfnamefont {S.~P.}\ \bibnamefont {Sharma}},\ and\
  \bibinfo {author} {\bibfnamefont {T.}~\bibnamefont {Mishra}},\ }\bibfield
  {title} {\emph {\bibinfo {title} {Dynamics of interacting particles on a
  rhombus chain: {Aharonov}-{Bohm} caging and inverse {Anderson} transition}},\
  }\href@noop {} {\Eprint {https://arxiv.org/abs/2409.05853} {arXiv preprint,
  arXiv:2409.05853}  (\bibinfo {year} {2024})}\BibitemShut {NoStop}%
\bibitem [{\citenamefont {Nicolau}\ \emph {et~al.}(2023)\citenamefont
  {Nicolau}, \citenamefont {Marques}, \citenamefont {Dias}, \citenamefont
  {Mompart},\ and\ \citenamefont {Ahufinger}}]{Nicolau2023}%
  \BibitemOpen
  \bibfield  {author} {\bibinfo {author} {\bibfnamefont {E.}~\bibnamefont
  {Nicolau}}, \bibinfo {author} {\bibfnamefont {A.~M.}\ \bibnamefont
  {Marques}}, \bibinfo {author} {\bibfnamefont {R.~G.}\ \bibnamefont {Dias}},
  \bibinfo {author} {\bibfnamefont {J.}~\bibnamefont {Mompart}},\ and\ \bibinfo
  {author} {\bibfnamefont {V.}~\bibnamefont {Ahufinger}},\ }\bibfield  {title}
  {\emph {\bibinfo {title} {Many-body {Aharonov}-{Bohm} caging in a lattice of
  rings}},\ }\href {https://doi.org/10.1103/physreva.107.023305} {\bibfield
  {journal} {\bibinfo  {journal} {Phys. Rev. A}\ }\textbf {\bibinfo {volume}
  {107}},\ \bibinfo {pages} {023305} (\bibinfo {year} {2023})}\BibitemShut
  {NoStop}%
\bibitem [{\citenamefont {Vepsäläinen}\ \emph {et~al.}(2019)\citenamefont
  {Vepsäläinen}, \citenamefont {Danilin},\ and\ \citenamefont
  {Paraoanu}}]{Vepsaelaeinen2019}%
  \BibitemOpen
  \bibfield  {author} {\bibinfo {author} {\bibfnamefont {A.}~\bibnamefont
  {Vepsäläinen}}, \bibinfo {author} {\bibfnamefont {S.}~\bibnamefont
  {Danilin}},\ and\ \bibinfo {author} {\bibfnamefont {G.~S.}\ \bibnamefont
  {Paraoanu}},\ }\bibfield  {title} {\emph {\bibinfo {title} {Superadiabatic
  population transfer in a three-level superconducting circuit}},\ }\href
  {https://doi.org/10.1126/sciadv.aau5999} {\bibfield  {journal} {\bibinfo
  {journal} {Sci. Adv.}\ }\textbf {\bibinfo {volume} {5}},\ \bibinfo {pages}
  {2} (\bibinfo {year} {2019})}\BibitemShut {NoStop}%
\bibitem [{\citenamefont {Vepsäläinen}\ and\ \citenamefont
  {Paraoanu}(2020)}]{Vepsaelaeinen2020}%
  \BibitemOpen
  \bibfield  {author} {\bibinfo {author} {\bibfnamefont {A.}~\bibnamefont
  {Vepsäläinen}}\ and\ \bibinfo {author} {\bibfnamefont {G.~S.}\ \bibnamefont
  {Paraoanu}},\ }\bibfield  {title} {\emph {\bibinfo {title} {Simulating spin
  chains using a superconducting circuit: Gauge invariance, superadiabatic
  transport, and broken time‐reversal symmetry}},\ }\href
  {https://doi.org/10.1002/qute.201900121} {\bibfield  {journal} {\bibinfo
  {journal} {Adv. Quantum Technol.}\ }\textbf {\bibinfo {volume} {3}},\
  \bibinfo {pages} {4} (\bibinfo {year} {2020})}\BibitemShut {NoStop}%
\bibitem [{\citenamefont {Rosen}\ \emph
  {et~al.}(2024{\natexlab{a}})\citenamefont {Rosen}, \citenamefont
  {Muschinske}, \citenamefont {Barrett}, \citenamefont {Chatterjee},
  \citenamefont {Hays}, \citenamefont {DeMarco}, \citenamefont {Karamlou},
  \citenamefont {Rower}, \citenamefont {Das}, \citenamefont {Kim},
  \citenamefont {Niedzielski}, \citenamefont {Schuldt}, \citenamefont
  {Serniak}, \citenamefont {Schwartz}, \citenamefont {Yoder}, \citenamefont
  {Grover},\ and\ \citenamefont {Oliver}}]{Rosen2024a}%
  \BibitemOpen
  \bibfield  {author} {\bibinfo {author} {\bibfnamefont {I.~T.}\ \bibnamefont
  {Rosen}}, \bibinfo {author} {\bibfnamefont {S.}~\bibnamefont {Muschinske}},
  \bibinfo {author} {\bibfnamefont {C.~N.}\ \bibnamefont {Barrett}}, \bibinfo
  {author} {\bibfnamefont {A.}~\bibnamefont {Chatterjee}}, \bibinfo {author}
  {\bibfnamefont {M.}~\bibnamefont {Hays}}, \bibinfo {author} {\bibfnamefont
  {M.~A.}\ \bibnamefont {DeMarco}}, \bibinfo {author} {\bibfnamefont {A.~H.}\
  \bibnamefont {Karamlou}}, \bibinfo {author} {\bibfnamefont {D.~A.}\
  \bibnamefont {Rower}}, \bibinfo {author} {\bibfnamefont {R.}~\bibnamefont
  {Das}}, \bibinfo {author} {\bibfnamefont {D.~K.}\ \bibnamefont {Kim}},
  \bibinfo {author} {\bibfnamefont {B.~M.}\ \bibnamefont {Niedzielski}},
  \bibinfo {author} {\bibfnamefont {M.}~\bibnamefont {Schuldt}}, \bibinfo
  {author} {\bibfnamefont {K.}~\bibnamefont {Serniak}}, \bibinfo {author}
  {\bibfnamefont {M.~E.}\ \bibnamefont {Schwartz}}, \bibinfo {author}
  {\bibfnamefont {J.~L.}\ \bibnamefont {Yoder}}, \bibinfo {author}
  {\bibfnamefont {J.~A.}\ \bibnamefont {Grover}},\ and\ \bibinfo {author}
  {\bibfnamefont {W.~D.}\ \bibnamefont {Oliver}},\ }\bibfield  {title} {\emph
  {\bibinfo {title} {A synthetic magnetic vector potential in a 2d
  superconducting qubit array}},\ }\href
  {https://doi.org/10.1038/s41567-024-02661-3} {\bibfield  {journal} {\bibinfo
  {journal} {Nat. Phys.}\ }\textbf {\bibinfo {volume} {20}},\ \bibinfo {pages}
  {1881} (\bibinfo {year} {2024}{\natexlab{a}})}\BibitemShut {NoStop}%
\bibitem [{\citenamefont {Roushan}\ \emph {et~al.}(2016)\citenamefont
  {Roushan}, \citenamefont {Neill}, \citenamefont {Megrant}, \citenamefont
  {Chen}, \citenamefont {Babbush}, \citenamefont {Barends}, \citenamefont
  {Campbell}, \citenamefont {Chen}, \citenamefont {Chiaro}, \citenamefont
  {Dunsworth} \emph {et~al.}}]{Roushan2016}%
  \BibitemOpen
  \bibfield  {author} {\bibinfo {author} {\bibfnamefont {P.}~\bibnamefont
  {Roushan}}, \bibinfo {author} {\bibfnamefont {C.}~\bibnamefont {Neill}},
  \bibinfo {author} {\bibfnamefont {A.}~\bibnamefont {Megrant}}, \bibinfo
  {author} {\bibfnamefont {Y.}~\bibnamefont {Chen}}, \bibinfo {author}
  {\bibfnamefont {R.}~\bibnamefont {Babbush}}, \bibinfo {author} {\bibfnamefont
  {R.}~\bibnamefont {Barends}}, \bibinfo {author} {\bibfnamefont
  {B.}~\bibnamefont {Campbell}}, \bibinfo {author} {\bibfnamefont
  {Z.}~\bibnamefont {Chen}}, \bibinfo {author} {\bibfnamefont {B.}~\bibnamefont
  {Chiaro}}, \bibinfo {author} {\bibfnamefont {A.}~\bibnamefont {Dunsworth}},
  \emph {et~al.},\ }\bibfield  {title} {\emph {\bibinfo {title} {Chiral
  ground-state currents of interacting photons in a synthetic magnetic
  field}},\ }\href {https://doi.org/10.1038/nphys3930} {\bibfield  {journal}
  {\bibinfo  {journal} {Nat. Phys.}\ }\textbf {\bibinfo {volume} {13}},\
  \bibinfo {pages} {146} (\bibinfo {year} {2016})}\BibitemShut {NoStop}%
\bibitem [{\citenamefont {Wang}\ \emph {et~al.}(2019)\citenamefont {Wang},
  \citenamefont {Song}, \citenamefont {Feng}, \citenamefont {Cai},
  \citenamefont {Xu}, \citenamefont {Deng}, \citenamefont {Li}, \citenamefont
  {Zheng}, \citenamefont {Zhu}, \citenamefont {Wang}, \citenamefont {Zhu},\
  and\ \citenamefont {Scully}}]{Wang2019}%
  \BibitemOpen
  \bibfield  {author} {\bibinfo {author} {\bibfnamefont {D.-W.}\ \bibnamefont
  {Wang}}, \bibinfo {author} {\bibfnamefont {C.}~\bibnamefont {Song}}, \bibinfo
  {author} {\bibfnamefont {W.}~\bibnamefont {Feng}}, \bibinfo {author}
  {\bibfnamefont {H.}~\bibnamefont {Cai}}, \bibinfo {author} {\bibfnamefont
  {D.}~\bibnamefont {Xu}}, \bibinfo {author} {\bibfnamefont {H.}~\bibnamefont
  {Deng}}, \bibinfo {author} {\bibfnamefont {H.}~\bibnamefont {Li}}, \bibinfo
  {author} {\bibfnamefont {D.}~\bibnamefont {Zheng}}, \bibinfo {author}
  {\bibfnamefont {X.}~\bibnamefont {Zhu}}, \bibinfo {author} {\bibfnamefont
  {H.}~\bibnamefont {Wang}}, \bibinfo {author} {\bibfnamefont {S.-Y.}\
  \bibnamefont {Zhu}},\ and\ \bibinfo {author} {\bibfnamefont {M.~O.}\
  \bibnamefont {Scully}},\ }\bibfield  {title} {\emph {\bibinfo {title}
  {Synthesis of antisymmetric spin exchange interaction and chiral spin
  clusters in superconducting circuits}},\ }\href
  {https://doi.org/10.1038/s41567-018-0400-9} {\bibfield  {journal} {\bibinfo
  {journal} {Nat. Phys.}\ }\textbf {\bibinfo {volume} {15}},\ \bibinfo {pages}
  {382} (\bibinfo {year} {2019})}\BibitemShut {NoStop}%
\bibitem [{\citenamefont {Liu}\ \emph {et~al.}(2020)\citenamefont {Liu},
  \citenamefont {Feng}, \citenamefont {Ren}, \citenamefont {Wang},\ and\
  \citenamefont {Wang}}]{Liu2020}%
  \BibitemOpen
  \bibfield  {author} {\bibinfo {author} {\bibfnamefont {W.}~\bibnamefont
  {Liu}}, \bibinfo {author} {\bibfnamefont {W.}~\bibnamefont {Feng}}, \bibinfo
  {author} {\bibfnamefont {W.}~\bibnamefont {Ren}}, \bibinfo {author}
  {\bibfnamefont {D.-W.}\ \bibnamefont {Wang}},\ and\ \bibinfo {author}
  {\bibfnamefont {H.}~\bibnamefont {Wang}},\ }\bibfield  {title} {\emph
  {\bibinfo {title} {Synthesizing three-body interaction of spin chirality with
  superconducting qubits}},\ }\href {https://doi.org/10.1063/1.5140884}
  {\bibfield  {journal} {\bibinfo  {journal} {Appl. Phys. Lett.}\ }\textbf
  {\bibinfo {volume} {116}},\ \bibinfo {pages} {114001} (\bibinfo {year}
  {2020})}\BibitemShut {NoStop}%
\bibitem [{\citenamefont {Zhang}\ \emph {et~al.}(2024)\citenamefont {Zhang},
  \citenamefont {Huang}, \citenamefont {Chu}, \citenamefont {Qiu},
  \citenamefont {Sun}, \citenamefont {Tao}, \citenamefont {Zhang},
  \citenamefont {Zhang}, \citenamefont {Zhou}, \citenamefont {Chen},
  \citenamefont {Liu}, \citenamefont {Liu}, \citenamefont {Zhong},
  \citenamefont {Miao}, \citenamefont {Niu},\ and\ \citenamefont
  {Yu}}]{ZJJ2024}%
  \BibitemOpen
  \bibfield  {author} {\bibinfo {author} {\bibfnamefont {J.}~\bibnamefont
  {Zhang}}, \bibinfo {author} {\bibfnamefont {W.}~\bibnamefont {Huang}},
  \bibinfo {author} {\bibfnamefont {J.}~\bibnamefont {Chu}}, \bibinfo {author}
  {\bibfnamefont {J.}~\bibnamefont {Qiu}}, \bibinfo {author} {\bibfnamefont
  {X.}~\bibnamefont {Sun}}, \bibinfo {author} {\bibfnamefont {Z.}~\bibnamefont
  {Tao}}, \bibinfo {author} {\bibfnamefont {J.}~\bibnamefont {Zhang}}, \bibinfo
  {author} {\bibfnamefont {L.}~\bibnamefont {Zhang}}, \bibinfo {author}
  {\bibfnamefont {Y.}~\bibnamefont {Zhou}}, \bibinfo {author} {\bibfnamefont
  {Y.}~\bibnamefont {Chen}}, \bibinfo {author} {\bibfnamefont {Y.}~\bibnamefont
  {Liu}}, \bibinfo {author} {\bibfnamefont {S.}~\bibnamefont {Liu}}, \bibinfo
  {author} {\bibfnamefont {Y.}~\bibnamefont {Zhong}}, \bibinfo {author}
  {\bibfnamefont {J.-J.}\ \bibnamefont {Miao}}, \bibinfo {author}
  {\bibfnamefont {J.}~\bibnamefont {Niu}},\ and\ \bibinfo {author}
  {\bibfnamefont {D.}~\bibnamefont {Yu}},\ }\bibfield  {title} {\emph {\bibinfo
  {title} {Synthetic multi-dimensional {Aharonov}-{Bohm} cages in {Fock} state
  lattices}},\ }\href@noop {} {\Eprint {https://arxiv.org/abs/2412.09766}
  {arXiv preprint, arXiv:2412.09766}  (\bibinfo {year} {2024})}\BibitemShut
  {NoStop}%
\bibitem [{\citenamefont {Martinez}\ \emph {et~al.}(2023)\citenamefont
  {Martinez}, \citenamefont {Chiu}, \citenamefont {Smitham},\ and\
  \citenamefont {Houck}}]{Martinez2023}%
  \BibitemOpen
  \bibfield  {author} {\bibinfo {author} {\bibfnamefont {J.~G.~C.}\
  \bibnamefont {Martinez}}, \bibinfo {author} {\bibfnamefont {C.~S.}\
  \bibnamefont {Chiu}}, \bibinfo {author} {\bibfnamefont {B.~M.}\ \bibnamefont
  {Smitham}},\ and\ \bibinfo {author} {\bibfnamefont {A.~A.}\ \bibnamefont
  {Houck}},\ }\bibfield  {title} {\emph {\bibinfo {title} {Flat-band
  localization and interaction-induced delocalization of photons}},\ }\href
  {https://doi.org/10.1126/sciadv.adj7195} {\bibfield  {journal} {\bibinfo
  {journal} {Sci. Adv.}\ }\textbf {\bibinfo {volume} {9}},\ \bibinfo {pages}
  {50} (\bibinfo {year} {2023})}\BibitemShut {NoStop}%
\bibitem [{\citenamefont {Yan}\ \emph {et~al.}(2019)\citenamefont {Yan},
  \citenamefont {Zhang}, \citenamefont {Gong}, \citenamefont {Wu},
  \citenamefont {Zheng}, \citenamefont {Li}, \citenamefont {Wang},
  \citenamefont {Liang}, \citenamefont {Lin}, \citenamefont {Xu}, \citenamefont
  {Guo}, \citenamefont {Sun}, \citenamefont {Peng}, \citenamefont {Xia},
  \citenamefont {Deng}, \citenamefont {Rong}, \citenamefont {You},
  \citenamefont {Nori}, \citenamefont {Fan}, \citenamefont {Zhu},\ and\
  \citenamefont {Pan}}]{Yan2019}%
  \BibitemOpen
  \bibfield  {author} {\bibinfo {author} {\bibfnamefont {Z.}~\bibnamefont
  {Yan}}, \bibinfo {author} {\bibfnamefont {Y.-R.}\ \bibnamefont {Zhang}},
  \bibinfo {author} {\bibfnamefont {M.}~\bibnamefont {Gong}}, \bibinfo {author}
  {\bibfnamefont {Y.}~\bibnamefont {Wu}}, \bibinfo {author} {\bibfnamefont
  {Y.}~\bibnamefont {Zheng}}, \bibinfo {author} {\bibfnamefont
  {S.}~\bibnamefont {Li}}, \bibinfo {author} {\bibfnamefont {C.}~\bibnamefont
  {Wang}}, \bibinfo {author} {\bibfnamefont {F.}~\bibnamefont {Liang}},
  \bibinfo {author} {\bibfnamefont {J.}~\bibnamefont {Lin}}, \bibinfo {author}
  {\bibfnamefont {Y.}~\bibnamefont {Xu}}, \bibinfo {author} {\bibfnamefont
  {C.}~\bibnamefont {Guo}}, \bibinfo {author} {\bibfnamefont {L.}~\bibnamefont
  {Sun}}, \bibinfo {author} {\bibfnamefont {C.-Z.}\ \bibnamefont {Peng}},
  \bibinfo {author} {\bibfnamefont {K.}~\bibnamefont {Xia}}, \bibinfo {author}
  {\bibfnamefont {H.}~\bibnamefont {Deng}}, \bibinfo {author} {\bibfnamefont
  {H.}~\bibnamefont {Rong}}, \bibinfo {author} {\bibfnamefont {J.~Q.}\
  \bibnamefont {You}}, \bibinfo {author} {\bibfnamefont {F.}~\bibnamefont
  {Nori}}, \bibinfo {author} {\bibfnamefont {H.}~\bibnamefont {Fan}}, \bibinfo
  {author} {\bibfnamefont {X.}~\bibnamefont {Zhu}},\ and\ \bibinfo {author}
  {\bibfnamefont {J.-W.}\ \bibnamefont {Pan}},\ }\bibfield  {title} {\emph
  {\bibinfo {title} {Strongly correlated quantum walks with a 12-qubit
  superconducting processor}},\ }\href
  {https://doi.org/10.1126/science.aaw1611} {\bibfield  {journal} {\bibinfo
  {journal} {Science}\ }\textbf {\bibinfo {volume} {364}},\ \bibinfo {pages}
  {753} (\bibinfo {year} {2019})}\BibitemShut {NoStop}%
\bibitem [{\citenamefont {Ma}\ \emph {et~al.}(2019)\citenamefont {Ma},
  \citenamefont {Saxberg}, \citenamefont {Owens}, \citenamefont {Leung},
  \citenamefont {Lu}, \citenamefont {Simon},\ and\ \citenamefont
  {Schuster}}]{Ma2019}%
  \BibitemOpen
  \bibfield  {author} {\bibinfo {author} {\bibfnamefont {R.}~\bibnamefont
  {Ma}}, \bibinfo {author} {\bibfnamefont {B.}~\bibnamefont {Saxberg}},
  \bibinfo {author} {\bibfnamefont {C.}~\bibnamefont {Owens}}, \bibinfo
  {author} {\bibfnamefont {N.}~\bibnamefont {Leung}}, \bibinfo {author}
  {\bibfnamefont {Y.}~\bibnamefont {Lu}}, \bibinfo {author} {\bibfnamefont
  {J.}~\bibnamefont {Simon}},\ and\ \bibinfo {author} {\bibfnamefont {D.~I.}\
  \bibnamefont {Schuster}},\ }\bibfield  {title} {\emph {\bibinfo {title} {A
  dissipatively stabilized {Mott} insulator of photons}},\ }\href
  {https://doi.org/10.1038/s41586-019-0897-9} {\bibfield  {journal} {\bibinfo
  {journal} {Nature}\ }\textbf {\bibinfo {volume} {566}},\ \bibinfo {pages}
  {51} (\bibinfo {year} {2019})}\BibitemShut {NoStop}%
\bibitem [{\citenamefont {Deng}\ \emph {et~al.}(2022)\citenamefont {Deng},
  \citenamefont {Dong}, \citenamefont {Zhang}, \citenamefont {Wu},
  \citenamefont {Yuan}, \citenamefont {Zhu}, \citenamefont {Jin}, \citenamefont
  {Li}, \citenamefont {Wang}, \citenamefont {Cai}, \citenamefont {Song},
  \citenamefont {Wang}, \citenamefont {You},\ and\ \citenamefont
  {Wang}}]{Deng2022}%
  \BibitemOpen
  \bibfield  {author} {\bibinfo {author} {\bibfnamefont {J.}~\bibnamefont
  {Deng}}, \bibinfo {author} {\bibfnamefont {H.}~\bibnamefont {Dong}}, \bibinfo
  {author} {\bibfnamefont {C.}~\bibnamefont {Zhang}}, \bibinfo {author}
  {\bibfnamefont {Y.}~\bibnamefont {Wu}}, \bibinfo {author} {\bibfnamefont
  {J.}~\bibnamefont {Yuan}}, \bibinfo {author} {\bibfnamefont {X.}~\bibnamefont
  {Zhu}}, \bibinfo {author} {\bibfnamefont {F.}~\bibnamefont {Jin}}, \bibinfo
  {author} {\bibfnamefont {H.}~\bibnamefont {Li}}, \bibinfo {author}
  {\bibfnamefont {Z.}~\bibnamefont {Wang}}, \bibinfo {author} {\bibfnamefont
  {H.}~\bibnamefont {Cai}}, \bibinfo {author} {\bibfnamefont {C.}~\bibnamefont
  {Song}}, \bibinfo {author} {\bibfnamefont {H.}~\bibnamefont {Wang}}, \bibinfo
  {author} {\bibfnamefont {J.~Q.}\ \bibnamefont {You}},\ and\ \bibinfo {author}
  {\bibfnamefont {D.-W.}\ \bibnamefont {Wang}},\ }\bibfield  {title} {\emph
  {\bibinfo {title} {Observing the quantum topology of light}},\ }\href
  {https://doi.org/10.1126/science.ade6219} {\bibfield  {journal} {\bibinfo
  {journal} {Science}\ }\textbf {\bibinfo {volume} {378}},\ \bibinfo {pages}
  {966} (\bibinfo {year} {2022})}\BibitemShut {NoStop}%
\bibitem [{\citenamefont {Karamlou}\ \emph {et~al.}(2022)\citenamefont
  {Karamlou}, \citenamefont {Braumüller}, \citenamefont {Yanay}, \citenamefont
  {Di~Paolo}, \citenamefont {Harrington}, \citenamefont {Kannan}, \citenamefont
  {Kim}, \citenamefont {Kjaergaard}, \citenamefont {Melville}, \citenamefont
  {Muschinske} \emph {et~al.}}]{Karamlou2022}%
  \BibitemOpen
  \bibfield  {author} {\bibinfo {author} {\bibfnamefont {A.~H.}\ \bibnamefont
  {Karamlou}}, \bibinfo {author} {\bibfnamefont {J.}~\bibnamefont
  {Braumüller}}, \bibinfo {author} {\bibfnamefont {Y.}~\bibnamefont {Yanay}},
  \bibinfo {author} {\bibfnamefont {A.}~\bibnamefont {Di~Paolo}}, \bibinfo
  {author} {\bibfnamefont {P.~M.}\ \bibnamefont {Harrington}}, \bibinfo
  {author} {\bibfnamefont {B.}~\bibnamefont {Kannan}}, \bibinfo {author}
  {\bibfnamefont {D.}~\bibnamefont {Kim}}, \bibinfo {author} {\bibfnamefont
  {M.}~\bibnamefont {Kjaergaard}}, \bibinfo {author} {\bibfnamefont
  {A.}~\bibnamefont {Melville}}, \bibinfo {author} {\bibfnamefont
  {S.}~\bibnamefont {Muschinske}}, \emph {et~al.},\ }\bibfield  {title} {\emph
  {\bibinfo {title} {Quantum transport and localization in 1d and 2d
  tight-binding lattices}},\ }\href
  {https://doi.org/10.1038/s41534-022-00528-0} {\bibfield  {journal} {\bibinfo
  {journal} {npj Quantum Inf.}\ }\textbf {\bibinfo {volume} {8}},\ \bibinfo
  {pages} {35} (\bibinfo {year} {2022})}\BibitemShut {NoStop}%
\bibitem [{\citenamefont {Yao}\ \emph {et~al.}(2023)\citenamefont {Yao},
  \citenamefont {Xiang}, \citenamefont {Guo}, \citenamefont {Bao},
  \citenamefont {Yang}, \citenamefont {Song}, \citenamefont {Shi},
  \citenamefont {Zhu}, \citenamefont {Jin}, \citenamefont {Chen} \emph
  {et~al.}}]{Yao2023}%
  \BibitemOpen
  \bibfield  {author} {\bibinfo {author} {\bibfnamefont {Y.}~\bibnamefont
  {Yao}}, \bibinfo {author} {\bibfnamefont {L.}~\bibnamefont {Xiang}}, \bibinfo
  {author} {\bibfnamefont {Z.}~\bibnamefont {Guo}}, \bibinfo {author}
  {\bibfnamefont {Z.}~\bibnamefont {Bao}}, \bibinfo {author} {\bibfnamefont
  {Y.-F.}\ \bibnamefont {Yang}}, \bibinfo {author} {\bibfnamefont
  {Z.}~\bibnamefont {Song}}, \bibinfo {author} {\bibfnamefont {H.}~\bibnamefont
  {Shi}}, \bibinfo {author} {\bibfnamefont {X.}~\bibnamefont {Zhu}}, \bibinfo
  {author} {\bibfnamefont {F.}~\bibnamefont {Jin}}, \bibinfo {author}
  {\bibfnamefont {J.}~\bibnamefont {Chen}}, \emph {et~al.},\ }\bibfield
  {title} {\emph {\bibinfo {title} {Observation of many-body {Fock} space
  dynamics in two dimensions}},\ }\href
  {https://doi.org/10.1038/s41567-023-02133-0} {\bibfield  {journal} {\bibinfo
  {journal} {Nat. Phys.}\ }\textbf {\bibinfo {volume} {19}},\ \bibinfo {pages}
  {1459} (\bibinfo {year} {2023})}\BibitemShut {NoStop}%
\bibitem [{\citenamefont {Karamlou}\ \emph {et~al.}(2024)\citenamefont
  {Karamlou}, \citenamefont {Rosen}, \citenamefont {Muschinske}, \citenamefont
  {Barrett}, \citenamefont {Di~Paolo}, \citenamefont {Ding}, \citenamefont
  {Harrington}, \citenamefont {Hays}, \citenamefont {Das}, \citenamefont {Kim},
  \citenamefont {Niedzielski}, \citenamefont {Schuldt}, \citenamefont
  {Serniak}, \citenamefont {Schwartz}, \citenamefont {Yoder}, \citenamefont
  {Gustavsson}, \citenamefont {Yanay}, \citenamefont {Grover},\ and\
  \citenamefont {Oliver}}]{Karamlou2024}%
  \BibitemOpen
  \bibfield  {author} {\bibinfo {author} {\bibfnamefont {A.~H.}\ \bibnamefont
  {Karamlou}}, \bibinfo {author} {\bibfnamefont {I.~T.}\ \bibnamefont {Rosen}},
  \bibinfo {author} {\bibfnamefont {S.~E.}\ \bibnamefont {Muschinske}},
  \bibinfo {author} {\bibfnamefont {C.~N.}\ \bibnamefont {Barrett}}, \bibinfo
  {author} {\bibfnamefont {A.}~\bibnamefont {Di~Paolo}}, \bibinfo {author}
  {\bibfnamefont {L.}~\bibnamefont {Ding}}, \bibinfo {author} {\bibfnamefont
  {P.~M.}\ \bibnamefont {Harrington}}, \bibinfo {author} {\bibfnamefont
  {M.}~\bibnamefont {Hays}}, \bibinfo {author} {\bibfnamefont {R.}~\bibnamefont
  {Das}}, \bibinfo {author} {\bibfnamefont {D.~K.}\ \bibnamefont {Kim}},
  \bibinfo {author} {\bibfnamefont {B.~M.}\ \bibnamefont {Niedzielski}},
  \bibinfo {author} {\bibfnamefont {M.}~\bibnamefont {Schuldt}}, \bibinfo
  {author} {\bibfnamefont {K.}~\bibnamefont {Serniak}}, \bibinfo {author}
  {\bibfnamefont {M.~E.}\ \bibnamefont {Schwartz}}, \bibinfo {author}
  {\bibfnamefont {J.~L.}\ \bibnamefont {Yoder}}, \bibinfo {author}
  {\bibfnamefont {S.}~\bibnamefont {Gustavsson}}, \bibinfo {author}
  {\bibfnamefont {Y.}~\bibnamefont {Yanay}}, \bibinfo {author} {\bibfnamefont
  {J.~A.}\ \bibnamefont {Grover}},\ and\ \bibinfo {author} {\bibfnamefont
  {W.~D.}\ \bibnamefont {Oliver}},\ }\bibfield  {title} {\emph {\bibinfo
  {title} {Probing entanglement in a 2d hard-core {Bose}–{Hubbard}
  lattice}},\ }\href {https://doi.org/10.1038/s41586-024-07325-z} {\bibfield
  {journal} {\bibinfo  {journal} {Nature}\ }\textbf {\bibinfo {volume} {629}},\
  \bibinfo {pages} {561} (\bibinfo {year} {2024})}\BibitemShut {NoStop}%
\bibitem [{\citenamefont {Shi}\ \emph {et~al.}(2024)\citenamefont {Shi},
  \citenamefont {Sun}, \citenamefont {Wang}, \citenamefont {Wang},
  \citenamefont {Zhang}, \citenamefont {Ma}, \citenamefont {Liu}, \citenamefont
  {Zhao}, \citenamefont {Song}, \citenamefont {Liang}, \citenamefont {Mei},
  \citenamefont {Zhang}, \citenamefont {Li}, \citenamefont {Chen},
  \citenamefont {Song}, \citenamefont {Wang}, \citenamefont {Xue},
  \citenamefont {Yu}, \citenamefont {Huang}, \citenamefont {Xiang},
  \citenamefont {Xu}, \citenamefont {Zheng},\ and\ \citenamefont
  {Fan}}]{Shi2024}%
  \BibitemOpen
  \bibfield  {author} {\bibinfo {author} {\bibfnamefont {Y.-H.}\ \bibnamefont
  {Shi}}, \bibinfo {author} {\bibfnamefont {Z.-H.}\ \bibnamefont {Sun}},
  \bibinfo {author} {\bibfnamefont {Y.-Y.}\ \bibnamefont {Wang}}, \bibinfo
  {author} {\bibfnamefont {Z.-A.}\ \bibnamefont {Wang}}, \bibinfo {author}
  {\bibfnamefont {Y.-R.}\ \bibnamefont {Zhang}}, \bibinfo {author}
  {\bibfnamefont {W.-G.}\ \bibnamefont {Ma}}, \bibinfo {author} {\bibfnamefont
  {H.-T.}\ \bibnamefont {Liu}}, \bibinfo {author} {\bibfnamefont
  {K.}~\bibnamefont {Zhao}}, \bibinfo {author} {\bibfnamefont {J.-C.}\
  \bibnamefont {Song}}, \bibinfo {author} {\bibfnamefont {G.-H.}\ \bibnamefont
  {Liang}}, \bibinfo {author} {\bibfnamefont {Z.-Y.}\ \bibnamefont {Mei}},
  \bibinfo {author} {\bibfnamefont {J.-C.}\ \bibnamefont {Zhang}}, \bibinfo
  {author} {\bibfnamefont {H.}~\bibnamefont {Li}}, \bibinfo {author}
  {\bibfnamefont {C.-T.}\ \bibnamefont {Chen}}, \bibinfo {author}
  {\bibfnamefont {X.}~\bibnamefont {Song}}, \bibinfo {author} {\bibfnamefont
  {J.}~\bibnamefont {Wang}}, \bibinfo {author} {\bibfnamefont {G.}~\bibnamefont
  {Xue}}, \bibinfo {author} {\bibfnamefont {H.}~\bibnamefont {Yu}}, \bibinfo
  {author} {\bibfnamefont {K.}~\bibnamefont {Huang}}, \bibinfo {author}
  {\bibfnamefont {Z.}~\bibnamefont {Xiang}}, \bibinfo {author} {\bibfnamefont
  {K.}~\bibnamefont {Xu}}, \bibinfo {author} {\bibfnamefont {D.}~\bibnamefont
  {Zheng}},\ and\ \bibinfo {author} {\bibfnamefont {H.}~\bibnamefont {Fan}},\
  }\bibfield  {title} {\emph {\bibinfo {title} {Probing spin hydrodynamics on a
  superconducting quantum simulator}},\ }\href
  {https://doi.org/10.1038/s41467-024-52082-2} {\bibfield  {journal} {\bibinfo
  {journal} {Nat. Commun.}\ }\textbf {\bibinfo {volume} {15}},\ \bibinfo
  {pages} {7573} (\bibinfo {year} {2024})}\BibitemShut {NoStop}%
\bibitem [{\citenamefont {Xiang}\ \emph {et~al.}(2024)\citenamefont {Xiang},
  \citenamefont {Chen}, \citenamefont {Zhu}, \citenamefont {Song},
  \citenamefont {Bao}, \citenamefont {Zhu}, \citenamefont {Jin}, \citenamefont
  {Wang}, \citenamefont {Xu}, \citenamefont {Zou}, \citenamefont {Li},
  \citenamefont {Wang}, \citenamefont {Song}, \citenamefont {Yue},
  \citenamefont {Partridge}, \citenamefont {Guo}, \citenamefont {Mondaini},
  \citenamefont {Wang},\ and\ \citenamefont {Scalettar}}]{Xiang2024}%
  \BibitemOpen
  \bibfield  {author} {\bibinfo {author} {\bibfnamefont {L.}~\bibnamefont
  {Xiang}}, \bibinfo {author} {\bibfnamefont {J.}~\bibnamefont {Chen}},
  \bibinfo {author} {\bibfnamefont {Z.}~\bibnamefont {Zhu}}, \bibinfo {author}
  {\bibfnamefont {Z.}~\bibnamefont {Song}}, \bibinfo {author} {\bibfnamefont
  {Z.}~\bibnamefont {Bao}}, \bibinfo {author} {\bibfnamefont {X.}~\bibnamefont
  {Zhu}}, \bibinfo {author} {\bibfnamefont {F.}~\bibnamefont {Jin}}, \bibinfo
  {author} {\bibfnamefont {K.}~\bibnamefont {Wang}}, \bibinfo {author}
  {\bibfnamefont {S.}~\bibnamefont {Xu}}, \bibinfo {author} {\bibfnamefont
  {Y.}~\bibnamefont {Zou}}, \bibinfo {author} {\bibfnamefont {H.}~\bibnamefont
  {Li}}, \bibinfo {author} {\bibfnamefont {Z.}~\bibnamefont {Wang}}, \bibinfo
  {author} {\bibfnamefont {C.}~\bibnamefont {Song}}, \bibinfo {author}
  {\bibfnamefont {A.}~\bibnamefont {Yue}}, \bibinfo {author} {\bibfnamefont
  {J.}~\bibnamefont {Partridge}}, \bibinfo {author} {\bibfnamefont
  {Q.}~\bibnamefont {Guo}}, \bibinfo {author} {\bibfnamefont {R.}~\bibnamefont
  {Mondaini}}, \bibinfo {author} {\bibfnamefont {H.}~\bibnamefont {Wang}},\
  and\ \bibinfo {author} {\bibfnamefont {R.~T.}\ \bibnamefont {Scalettar}},\
  }\bibfield  {title} {\emph {\bibinfo {title} {Enhanced quantum state transfer
  by circumventing quantum chaotic behavior}},\ }\href
  {https://doi.org/10.1038/s41467-024-48791-3} {\bibfield  {journal} {\bibinfo
  {journal} {Nat. Commun.}\ }\textbf {\bibinfo {volume} {15}},\ \bibinfo
  {pages} {4918} (\bibinfo {year} {2024})}\BibitemShut {NoStop}%
\bibitem [{\citenamefont {Deng}\ \emph {et~al.}(2024)\citenamefont {Deng},
  \citenamefont {Liu}, \citenamefont {Zhang}, \citenamefont {Li}, \citenamefont
  {Liu}, \citenamefont {Chen}, \citenamefont {Liu}, \citenamefont {Lu},
  \citenamefont {Wang}, \citenamefont {Li}, \citenamefont {Fang}, \citenamefont
  {Zhou}, \citenamefont {Song}, \citenamefont {Xu}, \citenamefont {He},
  \citenamefont {Liu}, \citenamefont {Huang}, \citenamefont {Xiang},
  \citenamefont {Wang}, \citenamefont {Zheng}, \citenamefont {Xue},
  \citenamefont {Xu}, \citenamefont {Yu},\ and\ \citenamefont
  {Fan}}]{Deng2024}%
  \BibitemOpen
  \bibfield  {author} {\bibinfo {author} {\bibfnamefont {C.-L.}\ \bibnamefont
  {Deng}}, \bibinfo {author} {\bibfnamefont {Y.}~\bibnamefont {Liu}}, \bibinfo
  {author} {\bibfnamefont {Y.-R.}\ \bibnamefont {Zhang}}, \bibinfo {author}
  {\bibfnamefont {X.-G.}\ \bibnamefont {Li}}, \bibinfo {author} {\bibfnamefont
  {T.}~\bibnamefont {Liu}}, \bibinfo {author} {\bibfnamefont {C.-T.}\
  \bibnamefont {Chen}}, \bibinfo {author} {\bibfnamefont {T.}~\bibnamefont
  {Liu}}, \bibinfo {author} {\bibfnamefont {C.-W.}\ \bibnamefont {Lu}},
  \bibinfo {author} {\bibfnamefont {Y.-Y.}\ \bibnamefont {Wang}}, \bibinfo
  {author} {\bibfnamefont {T.-M.}\ \bibnamefont {Li}}, \bibinfo {author}
  {\bibfnamefont {C.-P.}\ \bibnamefont {Fang}}, \bibinfo {author}
  {\bibfnamefont {S.-Y.}\ \bibnamefont {Zhou}}, \bibinfo {author}
  {\bibfnamefont {J.-C.}\ \bibnamefont {Song}}, \bibinfo {author}
  {\bibfnamefont {Y.-S.}\ \bibnamefont {Xu}}, \bibinfo {author} {\bibfnamefont
  {Y.}~\bibnamefont {He}}, \bibinfo {author} {\bibfnamefont {Z.-H.}\
  \bibnamefont {Liu}}, \bibinfo {author} {\bibfnamefont {K.-X.}\ \bibnamefont
  {Huang}}, \bibinfo {author} {\bibfnamefont {Z.-C.}\ \bibnamefont {Xiang}},
  \bibinfo {author} {\bibfnamefont {J.-C.}\ \bibnamefont {Wang}}, \bibinfo
  {author} {\bibfnamefont {D.-N.}\ \bibnamefont {Zheng}}, \bibinfo {author}
  {\bibfnamefont {G.-M.}\ \bibnamefont {Xue}}, \bibinfo {author} {\bibfnamefont
  {K.}~\bibnamefont {Xu}}, \bibinfo {author} {\bibfnamefont {H.-F.}\
  \bibnamefont {Yu}},\ and\ \bibinfo {author} {\bibfnamefont {H.}~\bibnamefont
  {Fan}},\ }\bibfield  {title} {\emph {\bibinfo {title} {High-order topological
  pumping on a superconducting quantum processor}},\ }\href
  {https://doi.org/10.1103/physrevlett.133.140402} {\bibfield  {journal}
  {\bibinfo  {journal} {Phys. Rev. Lett.}\ }\textbf {\bibinfo {volume} {133}},\
  \bibinfo {pages} {140402} (\bibinfo {year} {2024})}\BibitemShut {NoStop}%
\bibitem [{\citenamefont {Liu}\ \emph {et~al.}(2025)\citenamefont {Liu},
  \citenamefont {Zhang}, \citenamefont {Shi}, \citenamefont {Liu},
  \citenamefont {Lu}, \citenamefont {Wang}, \citenamefont {Li}, \citenamefont
  {Li}, \citenamefont {Deng}, \citenamefont {Zhou}, \citenamefont {Liu},
  \citenamefont {Zhang}, \citenamefont {Liang}, \citenamefont {Mei},
  \citenamefont {Ma}, \citenamefont {Liu}, \citenamefont {Liu}, \citenamefont
  {Chen}, \citenamefont {Huang}, \citenamefont {Song}, \citenamefont {Zhao},
  \citenamefont {Tian}, \citenamefont {Xiang}, \citenamefont {Zheng},
  \citenamefont {Nori}, \citenamefont {Xu},\ and\ \citenamefont
  {Fan}}]{Liu2025}%
  \BibitemOpen
  \bibfield  {author} {\bibinfo {author} {\bibfnamefont {Y.}~\bibnamefont
  {Liu}}, \bibinfo {author} {\bibfnamefont {Y.-R.}\ \bibnamefont {Zhang}},
  \bibinfo {author} {\bibfnamefont {Y.-H.}\ \bibnamefont {Shi}}, \bibinfo
  {author} {\bibfnamefont {T.}~\bibnamefont {Liu}}, \bibinfo {author}
  {\bibfnamefont {C.}~\bibnamefont {Lu}}, \bibinfo {author} {\bibfnamefont
  {Y.-Y.}\ \bibnamefont {Wang}}, \bibinfo {author} {\bibfnamefont
  {H.}~\bibnamefont {Li}}, \bibinfo {author} {\bibfnamefont {T.-M.}\
  \bibnamefont {Li}}, \bibinfo {author} {\bibfnamefont {C.-L.}\ \bibnamefont
  {Deng}}, \bibinfo {author} {\bibfnamefont {S.-Y.}\ \bibnamefont {Zhou}},
  \bibinfo {author} {\bibfnamefont {T.}~\bibnamefont {Liu}}, \bibinfo {author}
  {\bibfnamefont {J.-C.}\ \bibnamefont {Zhang}}, \bibinfo {author}
  {\bibfnamefont {G.-H.}\ \bibnamefont {Liang}}, \bibinfo {author}
  {\bibfnamefont {Z.-Y.}\ \bibnamefont {Mei}}, \bibinfo {author} {\bibfnamefont
  {W.-G.}\ \bibnamefont {Ma}}, \bibinfo {author} {\bibfnamefont {H.-T.}\
  \bibnamefont {Liu}}, \bibinfo {author} {\bibfnamefont {Z.-H.}\ \bibnamefont
  {Liu}}, \bibinfo {author} {\bibfnamefont {C.-T.}\ \bibnamefont {Chen}},
  \bibinfo {author} {\bibfnamefont {K.}~\bibnamefont {Huang}}, \bibinfo
  {author} {\bibfnamefont {X.}~\bibnamefont {Song}}, \bibinfo {author}
  {\bibfnamefont {S.~P.}\ \bibnamefont {Zhao}}, \bibinfo {author}
  {\bibfnamefont {Y.}~\bibnamefont {Tian}}, \bibinfo {author} {\bibfnamefont
  {Z.}~\bibnamefont {Xiang}}, \bibinfo {author} {\bibfnamefont
  {D.}~\bibnamefont {Zheng}}, \bibinfo {author} {\bibfnamefont
  {F.}~\bibnamefont {Nori}}, \bibinfo {author} {\bibfnamefont {K.}~\bibnamefont
  {Xu}},\ and\ \bibinfo {author} {\bibfnamefont {H.}~\bibnamefont {Fan}},\
  }\bibfield  {title} {\emph {\bibinfo {title} {Interplay between disorder and
  topology in {Thouless} pumping on a superconducting quantum processor}},\
  }\href {https://doi.org/10.1038/s41467-024-55343-2} {\bibfield  {journal}
  {\bibinfo  {journal} {Nat. Commun.}\ }\textbf {\bibinfo {volume} {16}},\
  \bibinfo {pages} {108} (\bibinfo {year} {2025})}\BibitemShut {NoStop}%
\bibitem [{\citenamefont {Wang}\ \emph {et~al.}(2024)\citenamefont {Wang},
  \citenamefont {Liu}, \citenamefont {Chen}, \citenamefont {Chen},
  \citenamefont {Zhao}, \citenamefont {Ying}, \citenamefont {Shang},
  \citenamefont {Wang}, \citenamefont {Huo}, \citenamefont {Peng},
  \citenamefont {Zhu}, \citenamefont {Lu},\ and\ \citenamefont
  {Pan}}]{Wang2024Science}%
  \BibitemOpen
  \bibfield  {author} {\bibinfo {author} {\bibfnamefont {C.}~\bibnamefont
  {Wang}}, \bibinfo {author} {\bibfnamefont {F.-M.}\ \bibnamefont {Liu}},
  \bibinfo {author} {\bibfnamefont {M.-C.}\ \bibnamefont {Chen}}, \bibinfo
  {author} {\bibfnamefont {H.}~\bibnamefont {Chen}}, \bibinfo {author}
  {\bibfnamefont {X.-H.}\ \bibnamefont {Zhao}}, \bibinfo {author}
  {\bibfnamefont {C.}~\bibnamefont {Ying}}, \bibinfo {author} {\bibfnamefont
  {Z.-X.}\ \bibnamefont {Shang}}, \bibinfo {author} {\bibfnamefont {J.-W.}\
  \bibnamefont {Wang}}, \bibinfo {author} {\bibfnamefont {Y.-H.}\ \bibnamefont
  {Huo}}, \bibinfo {author} {\bibfnamefont {C.-Z.}\ \bibnamefont {Peng}},
  \bibinfo {author} {\bibfnamefont {X.}~\bibnamefont {Zhu}}, \bibinfo {author}
  {\bibfnamefont {C.-Y.}\ \bibnamefont {Lu}},\ and\ \bibinfo {author}
  {\bibfnamefont {J.-W.}\ \bibnamefont {Pan}},\ }\bibfield  {title} {\emph
  {\bibinfo {title} {Realization of fractional quantum {Hall} state with
  interacting photons}},\ }\href {https://doi.org/10.1126/science.ado3912}
  {\bibfield  {journal} {\bibinfo  {journal} {Science}\ }\textbf {\bibinfo
  {volume} {384}},\ \bibinfo {pages} {579} (\bibinfo {year}
  {2024})}\BibitemShut {NoStop}%
\bibitem [{\citenamefont {Neill}\ \emph {et~al.}(2021)\citenamefont {Neill},
  \citenamefont {McCourt}, \citenamefont {Mi}, \citenamefont {Jiang},
  \citenamefont {Niu}, \citenamefont {Mruczkiewicz}, \citenamefont {Aleiner},
  \citenamefont {Arute}, \citenamefont {Arya}, \citenamefont {Atalaya} \emph
  {et~al.}}]{Neill2021}%
  \BibitemOpen
  \bibfield  {author} {\bibinfo {author} {\bibfnamefont {C.}~\bibnamefont
  {Neill}}, \bibinfo {author} {\bibfnamefont {T.}~\bibnamefont {McCourt}},
  \bibinfo {author} {\bibfnamefont {X.}~\bibnamefont {Mi}}, \bibinfo {author}
  {\bibfnamefont {Z.}~\bibnamefont {Jiang}}, \bibinfo {author} {\bibfnamefont
  {M.~Y.}\ \bibnamefont {Niu}}, \bibinfo {author} {\bibfnamefont
  {W.}~\bibnamefont {Mruczkiewicz}}, \bibinfo {author} {\bibfnamefont
  {I.}~\bibnamefont {Aleiner}}, \bibinfo {author} {\bibfnamefont
  {F.}~\bibnamefont {Arute}}, \bibinfo {author} {\bibfnamefont
  {K.}~\bibnamefont {Arya}}, \bibinfo {author} {\bibfnamefont {J.}~\bibnamefont
  {Atalaya}}, \emph {et~al.},\ }\bibfield  {title} {\emph {\bibinfo {title}
  {Accurately computing the electronic properties of a quantum ring}},\ }\href
  {https://doi.org/10.1038/s41586-021-03576-2} {\bibfield  {journal} {\bibinfo
  {journal} {Nature}\ }\textbf {\bibinfo {volume} {594}},\ \bibinfo {pages}
  {508} (\bibinfo {year} {2021})}\BibitemShut {NoStop}%
\bibitem [{\citenamefont {Rosen}\ \emph
  {et~al.}(2024{\natexlab{b}})\citenamefont {Rosen}, \citenamefont
  {Muschinske}, \citenamefont {Barrett}, \citenamefont {Rower}, \citenamefont
  {Das}, \citenamefont {Kim}, \citenamefont {Niedzielski}, \citenamefont
  {Schuldt}, \citenamefont {Serniak}, \citenamefont {Schwartz}, \citenamefont
  {Yoder}, \citenamefont {Grover},\ and\ \citenamefont {Oliver}}]{Rosen2024}%
  \BibitemOpen
  \bibfield  {author} {\bibinfo {author} {\bibfnamefont {I.~T.}\ \bibnamefont
  {Rosen}}, \bibinfo {author} {\bibfnamefont {S.}~\bibnamefont {Muschinske}},
  \bibinfo {author} {\bibfnamefont {C.~N.}\ \bibnamefont {Barrett}}, \bibinfo
  {author} {\bibfnamefont {D.~A.}\ \bibnamefont {Rower}}, \bibinfo {author}
  {\bibfnamefont {R.}~\bibnamefont {Das}}, \bibinfo {author} {\bibfnamefont
  {D.~K.}\ \bibnamefont {Kim}}, \bibinfo {author} {\bibfnamefont {B.~M.}\
  \bibnamefont {Niedzielski}}, \bibinfo {author} {\bibfnamefont
  {M.}~\bibnamefont {Schuldt}}, \bibinfo {author} {\bibfnamefont
  {K.}~\bibnamefont {Serniak}}, \bibinfo {author} {\bibfnamefont {M.~E.}\
  \bibnamefont {Schwartz}}, \bibinfo {author} {\bibfnamefont {J.~L.}\
  \bibnamefont {Yoder}}, \bibinfo {author} {\bibfnamefont {J.~A.}\ \bibnamefont
  {Grover}},\ and\ \bibinfo {author} {\bibfnamefont {W.~D.}\ \bibnamefont
  {Oliver}},\ }\bibfield  {title} {\emph {\bibinfo {title} {Flat-band
  (de)localization emulated with a superconducting qubit array}},\ }\href@noop
  {} {\Eprint {https://arxiv.org/abs/2410.07878} {arXiv preprint,
  arXiv:2410.07878}  (\bibinfo {year} {2024}{\natexlab{b}})}\BibitemShut
  {NoStop}%
\bibitem [{\citenamefont {Yan}\ \emph {et~al.}(2018)\citenamefont {Yan},
  \citenamefont {Krantz}, \citenamefont {Sung}, \citenamefont {Kjaergaard},
  \citenamefont {Campbell}, \citenamefont {Orlando}, \citenamefont
  {Gustavsson},\ and\ \citenamefont {Oliver}}]{Yan2018}%
  \BibitemOpen
  \bibfield  {author} {\bibinfo {author} {\bibfnamefont {F.}~\bibnamefont
  {Yan}}, \bibinfo {author} {\bibfnamefont {P.}~\bibnamefont {Krantz}},
  \bibinfo {author} {\bibfnamefont {Y.}~\bibnamefont {Sung}}, \bibinfo {author}
  {\bibfnamefont {M.}~\bibnamefont {Kjaergaard}}, \bibinfo {author}
  {\bibfnamefont {D.~L.}\ \bibnamefont {Campbell}}, \bibinfo {author}
  {\bibfnamefont {T.~P.}\ \bibnamefont {Orlando}}, \bibinfo {author}
  {\bibfnamefont {S.}~\bibnamefont {Gustavsson}},\ and\ \bibinfo {author}
  {\bibfnamefont {W.~D.}\ \bibnamefont {Oliver}},\ }\bibfield  {title} {\emph
  {\bibinfo {title} {Tunable coupling scheme for implementing high-fidelity
  two-qubit gates}},\ }\href {https://doi.org/10.1103/PhysRevApplied.10.054062}
  {\bibfield  {journal} {\bibinfo  {journal} {Phys. Rev. Applied}\ }\textbf
  {\bibinfo {volume} {10}},\ \bibinfo {pages} {054062} (\bibinfo {year}
  {2018})}\BibitemShut {NoStop}%
\bibitem [{\citenamefont {Xu}\ \emph {et~al.}(2020)\citenamefont {Xu},
  \citenamefont {Chu}, \citenamefont {Yuan}, \citenamefont {Qiu}, \citenamefont
  {Zhou}, \citenamefont {Zhang}, \citenamefont {Tan}, \citenamefont {Yu},
  \citenamefont {Liu}, \citenamefont {Li} \emph {et~al.}}]{Xu2020}%
  \BibitemOpen
  \bibfield  {author} {\bibinfo {author} {\bibfnamefont {Y.}~\bibnamefont
  {Xu}}, \bibinfo {author} {\bibfnamefont {J.}~\bibnamefont {Chu}}, \bibinfo
  {author} {\bibfnamefont {J.}~\bibnamefont {Yuan}}, \bibinfo {author}
  {\bibfnamefont {J.}~\bibnamefont {Qiu}}, \bibinfo {author} {\bibfnamefont
  {Y.}~\bibnamefont {Zhou}}, \bibinfo {author} {\bibfnamefont {L.}~\bibnamefont
  {Zhang}}, \bibinfo {author} {\bibfnamefont {X.}~\bibnamefont {Tan}}, \bibinfo
  {author} {\bibfnamefont {Y.}~\bibnamefont {Yu}}, \bibinfo {author}
  {\bibfnamefont {S.}~\bibnamefont {Liu}}, \bibinfo {author} {\bibfnamefont
  {J.}~\bibnamefont {Li}}, \emph {et~al.},\ }\bibfield  {title} {\emph
  {\bibinfo {title} {High-fidelity, high-scalability two-qubit gate scheme for
  superconducting qubits}},\ }\href
  {https://doi.org/10.1103/PhysRevLett.125.240503} {\bibfield  {journal}
  {\bibinfo  {journal} {Phys. Rev. Lett.}\ }\textbf {\bibinfo {volume} {125}},\
  \bibinfo {pages} {240503} (\bibinfo {year} {2020})}\BibitemShut {NoStop}%
\bibitem [{\citenamefont {Martinez~Alvarez}\ and\ \citenamefont
  {Coutinho-Filho}(2019)}]{MartinezAlvarez2019}%
  \BibitemOpen
  \bibfield  {author} {\bibinfo {author} {\bibfnamefont {V.~M.}\ \bibnamefont
  {Martinez~Alvarez}}\ and\ \bibinfo {author} {\bibfnamefont {M.~D.}\
  \bibnamefont {Coutinho-Filho}},\ }\bibfield  {title} {\emph {\bibinfo {title}
  {Edge states in trimer lattices}},\ }\href
  {https://doi.org/10.1103/physreva.99.013833} {\bibfield  {journal} {\bibinfo
  {journal} {Phys. Rev. A}\ }\textbf {\bibinfo {volume} {99}},\ \bibinfo
  {pages} {013833} (\bibinfo {year} {2019})}\BibitemShut {NoStop}%
\bibitem [{\citenamefont {Anastasiadis}\ \emph {et~al.}(2022)\citenamefont
  {Anastasiadis}, \citenamefont {Styliaris}, \citenamefont {Chaunsali},
  \citenamefont {Theocharis},\ and\ \citenamefont
  {Diakonos}}]{Anastasiadis2022}%
  \BibitemOpen
  \bibfield  {author} {\bibinfo {author} {\bibfnamefont {A.}~\bibnamefont
  {Anastasiadis}}, \bibinfo {author} {\bibfnamefont {G.}~\bibnamefont
  {Styliaris}}, \bibinfo {author} {\bibfnamefont {R.}~\bibnamefont
  {Chaunsali}}, \bibinfo {author} {\bibfnamefont {G.}~\bibnamefont
  {Theocharis}},\ and\ \bibinfo {author} {\bibfnamefont {F.~K.}\ \bibnamefont
  {Diakonos}},\ }\bibfield  {title} {\emph {\bibinfo {title} {Bulk-edge
  correspondence in the trimer {Su-Schrieffer-Heeger} model}},\ }\href
  {https://doi.org/10.1103/physrevb.106.085109} {\bibfield  {journal} {\bibinfo
   {journal} {Phys. Rev. B}\ }\textbf {\bibinfo {volume} {106}},\ \bibinfo
  {pages} {085109} (\bibinfo {year} {2022})}\BibitemShut {NoStop}%
\bibitem [{\citenamefont {Zhou}\ \emph {et~al.}(2023)\citenamefont {Zhou},
  \citenamefont {Wang}, \citenamefont {Poon}, \citenamefont {Zhou},\ and\
  \citenamefont {Liu}}]{Zhou2023}%
  \BibitemOpen
  \bibfield  {author} {\bibinfo {author} {\bibfnamefont {X.-C.}\ \bibnamefont
  {Zhou}}, \bibinfo {author} {\bibfnamefont {Y.}~\bibnamefont {Wang}}, \bibinfo
  {author} {\bibfnamefont {T.-F.~J.}\ \bibnamefont {Poon}}, \bibinfo {author}
  {\bibfnamefont {Q.}~\bibnamefont {Zhou}},\ and\ \bibinfo {author}
  {\bibfnamefont {X.-J.}\ \bibnamefont {Liu}},\ }\bibfield  {title} {\emph
  {\bibinfo {title} {Exact new mobility edges between critical and localized
  states}},\ }\href {https://doi.org/10.1103/physrevlett.131.176401} {\bibfield
   {journal} {\bibinfo  {journal} {Phys. Rev. Lett.}\ }\textbf {\bibinfo
  {volume} {131}},\ \bibinfo {pages} {176401} (\bibinfo {year}
  {2023})}\BibitemShut {NoStop}%
\bibitem [{\citenamefont {Danieli}\ \emph
  {et~al.}(2021{\natexlab{b}})\citenamefont {Danieli}, \citenamefont
  {Andreanov}, \citenamefont {Mithun},\ and\ \citenamefont
  {Flach}}]{Danieli2021}%
  \BibitemOpen
  \bibfield  {author} {\bibinfo {author} {\bibfnamefont {C.}~\bibnamefont
  {Danieli}}, \bibinfo {author} {\bibfnamefont {A.}~\bibnamefont {Andreanov}},
  \bibinfo {author} {\bibfnamefont {T.}~\bibnamefont {Mithun}},\ and\ \bibinfo
  {author} {\bibfnamefont {S.}~\bibnamefont {Flach}},\ }\bibfield  {title}
  {\emph {\bibinfo {title} {Nonlinear caging in all-bands-flat lattices}},\
  }\href {https://doi.org/10.1103/physrevb.104.085131} {\bibfield  {journal}
  {\bibinfo  {journal} {Phys. Rev. B}\ }\textbf {\bibinfo {volume} {104}},\
  \bibinfo {pages} {085131} (\bibinfo {year} {2021}{\natexlab{b}})}\BibitemShut
  {NoStop}%
\bibitem [{\citenamefont {Paladugu}\ \emph {et~al.}(2024)\citenamefont
  {Paladugu}, \citenamefont {Chen}, \citenamefont {An}, \citenamefont {Yan},\
  and\ \citenamefont {Gadway}}]{Paladugu2024}%
  \BibitemOpen
  \bibfield  {author} {\bibinfo {author} {\bibfnamefont {S.~N.~M.}\
  \bibnamefont {Paladugu}}, \bibinfo {author} {\bibfnamefont {T.}~\bibnamefont
  {Chen}}, \bibinfo {author} {\bibfnamefont {F.~A.}\ \bibnamefont {An}},
  \bibinfo {author} {\bibfnamefont {B.}~\bibnamefont {Yan}},\ and\ \bibinfo
  {author} {\bibfnamefont {B.}~\bibnamefont {Gadway}},\ }\bibfield  {title}
  {\emph {\bibinfo {title} {Injection spectroscopy of momentum state
  lattices}},\ }\href {https://doi.org/10.1038/s42005-024-01526-8} {\bibfield
  {journal} {\bibinfo  {journal} {Commun. Phys.}\ }\textbf {\bibinfo {volume}
  {7}},\ \bibinfo {pages} {39} (\bibinfo {year} {2024})}\BibitemShut {NoStop}%
\bibitem [{\citenamefont {Saxberg}\ \emph {et~al.}(2022)\citenamefont
  {Saxberg}, \citenamefont {Vrajitoarea}, \citenamefont {Roberts},
  \citenamefont {Panetta}, \citenamefont {Simon},\ and\ \citenamefont
  {Schuster}}]{Saxberg2022}%
  \BibitemOpen
  \bibfield  {author} {\bibinfo {author} {\bibfnamefont {B.}~\bibnamefont
  {Saxberg}}, \bibinfo {author} {\bibfnamefont {A.}~\bibnamefont
  {Vrajitoarea}}, \bibinfo {author} {\bibfnamefont {G.}~\bibnamefont
  {Roberts}}, \bibinfo {author} {\bibfnamefont {M.~G.}\ \bibnamefont
  {Panetta}}, \bibinfo {author} {\bibfnamefont {J.}~\bibnamefont {Simon}},\
  and\ \bibinfo {author} {\bibfnamefont {D.~I.}\ \bibnamefont {Schuster}},\
  }\bibfield  {title} {\emph {\bibinfo {title} {Disorder-assisted assembly of
  strongly correlated fluids of light}},\ }\href
  {https://doi.org/10.1038/s41586-022-05357-x} {\bibfield  {journal} {\bibinfo
  {journal} {Nature}\ }\textbf {\bibinfo {volume} {612}},\ \bibinfo {pages}
  {435} (\bibinfo {year} {2022})}\BibitemShut {NoStop}%
\bibitem [{\citenamefont {Shi}\ \emph {et~al.}(2023)\citenamefont {Shi},
  \citenamefont {Yang}, \citenamefont {Xiang}, \citenamefont {Ge},
  \citenamefont {Li}, \citenamefont {Wang}, \citenamefont {Huang},
  \citenamefont {Tian}, \citenamefont {Song}, \citenamefont {Zheng},
  \citenamefont {Xu}, \citenamefont {Cai},\ and\ \citenamefont
  {Fan}}]{Shi2023}%
  \BibitemOpen
  \bibfield  {author} {\bibinfo {author} {\bibfnamefont {Y.-H.}\ \bibnamefont
  {Shi}}, \bibinfo {author} {\bibfnamefont {R.-Q.}\ \bibnamefont {Yang}},
  \bibinfo {author} {\bibfnamefont {Z.}~\bibnamefont {Xiang}}, \bibinfo
  {author} {\bibfnamefont {Z.-Y.}\ \bibnamefont {Ge}}, \bibinfo {author}
  {\bibfnamefont {H.}~\bibnamefont {Li}}, \bibinfo {author} {\bibfnamefont
  {Y.-Y.}\ \bibnamefont {Wang}}, \bibinfo {author} {\bibfnamefont
  {K.}~\bibnamefont {Huang}}, \bibinfo {author} {\bibfnamefont
  {Y.}~\bibnamefont {Tian}}, \bibinfo {author} {\bibfnamefont {X.}~\bibnamefont
  {Song}}, \bibinfo {author} {\bibfnamefont {D.}~\bibnamefont {Zheng}},
  \bibinfo {author} {\bibfnamefont {K.}~\bibnamefont {Xu}}, \bibinfo {author}
  {\bibfnamefont {R.-G.}\ \bibnamefont {Cai}},\ and\ \bibinfo {author}
  {\bibfnamefont {H.}~\bibnamefont {Fan}},\ }\bibfield  {title} {\emph
  {\bibinfo {title} {Quantum simulation of hawking radiation and curved
  spacetime with a superconducting on-chip black hole}},\ }\href
  {https://doi.org/10.1038/s41467-023-39064-6} {\bibfield  {journal} {\bibinfo
  {journal} {Nat. Commun.}\ }\textbf {\bibinfo {volume} {14}},\ \bibinfo
  {pages} {3263} (\bibinfo {year} {2023})}\BibitemShut {NoStop}%
\bibitem [{\citenamefont {Röntgen}\ \emph {et~al.}(2019)\citenamefont
  {Röntgen}, \citenamefont {Morfonios}, \citenamefont {Brouzos}, \citenamefont
  {Diakonos},\ and\ \citenamefont {Schmelcher}}]{Roentgen2019}%
  \BibitemOpen
  \bibfield  {author} {\bibinfo {author} {\bibfnamefont {M.}~\bibnamefont
  {Röntgen}}, \bibinfo {author} {\bibfnamefont {C.}~\bibnamefont {Morfonios}},
  \bibinfo {author} {\bibfnamefont {I.}~\bibnamefont {Brouzos}}, \bibinfo
  {author} {\bibfnamefont {F.}~\bibnamefont {Diakonos}},\ and\ \bibinfo
  {author} {\bibfnamefont {P.}~\bibnamefont {Schmelcher}},\ }\bibfield  {title}
  {\emph {\bibinfo {title} {Quantum network transfer and storage with compact
  localized states induced by local symmetries}},\ }\href
  {https://doi.org/10.1103/physrevlett.123.080504} {\bibfield  {journal}
  {\bibinfo  {journal} {Phys. Rev. Lett.}\ }\textbf {\bibinfo {volume} {123}},\
  \bibinfo {pages} {080504} (\bibinfo {year} {2019})}\BibitemShut {NoStop}%
\bibitem [{\citenamefont {Campbell}\ \emph {et~al.}(2020)\citenamefont
  {Campbell}, \citenamefont {Shim}, \citenamefont {Kannan}, \citenamefont
  {Winik}, \citenamefont {Kim}, \citenamefont {Melville}, \citenamefont
  {Niedzielski}, \citenamefont {Yoder}, \citenamefont {Tahan}, \citenamefont
  {Gustavsson},\ and\ \citenamefont {Oliver}}]{Campbell2020}%
  \BibitemOpen
  \bibfield  {author} {\bibinfo {author} {\bibfnamefont {D.~L.}\ \bibnamefont
  {Campbell}}, \bibinfo {author} {\bibfnamefont {Y.-P.}\ \bibnamefont {Shim}},
  \bibinfo {author} {\bibfnamefont {B.}~\bibnamefont {Kannan}}, \bibinfo
  {author} {\bibfnamefont {R.}~\bibnamefont {Winik}}, \bibinfo {author}
  {\bibfnamefont {D.~K.}\ \bibnamefont {Kim}}, \bibinfo {author} {\bibfnamefont
  {A.}~\bibnamefont {Melville}}, \bibinfo {author} {\bibfnamefont {B.~M.}\
  \bibnamefont {Niedzielski}}, \bibinfo {author} {\bibfnamefont {J.~L.}\
  \bibnamefont {Yoder}}, \bibinfo {author} {\bibfnamefont {C.}~\bibnamefont
  {Tahan}}, \bibinfo {author} {\bibfnamefont {S.}~\bibnamefont {Gustavsson}},\
  and\ \bibinfo {author} {\bibfnamefont {W.~D.}\ \bibnamefont {Oliver}},\
  }\bibfield  {title} {\emph {\bibinfo {title} {Universal nonadiabatic control
  of small-gap superconducting qubits}},\ }\href
  {https://doi.org/10.1103/physrevx.10.041051} {\bibfield  {journal} {\bibinfo
  {journal} {Phys. Rev. X}\ }\textbf {\bibinfo {volume} {10}},\ \bibinfo
  {pages} {041051} (\bibinfo {year} {2020})}\BibitemShut {NoStop}%
\bibitem [{\citenamefont {Teoh}\ \emph {et~al.}(2023)\citenamefont {Teoh},
  \citenamefont {Winkel}, \citenamefont {Babla}, \citenamefont {Chapman},
  \citenamefont {Claes}, \citenamefont {de~Graaf}, \citenamefont {Garmon},
  \citenamefont {Kalfus}, \citenamefont {Lu}, \citenamefont {Maiti},
  \citenamefont {Sahay}, \citenamefont {Thakur}, \citenamefont {Tsunoda},
  \citenamefont {Xue}, \citenamefont {Frunzio}, \citenamefont {Girvin},
  \citenamefont {Puri},\ and\ \citenamefont {Schoelkopf}}]{Teoh2023}%
  \BibitemOpen
  \bibfield  {author} {\bibinfo {author} {\bibfnamefont {J.~D.}\ \bibnamefont
  {Teoh}}, \bibinfo {author} {\bibfnamefont {P.}~\bibnamefont {Winkel}},
  \bibinfo {author} {\bibfnamefont {H.~K.}\ \bibnamefont {Babla}}, \bibinfo
  {author} {\bibfnamefont {B.~J.}\ \bibnamefont {Chapman}}, \bibinfo {author}
  {\bibfnamefont {J.}~\bibnamefont {Claes}}, \bibinfo {author} {\bibfnamefont
  {S.~J.}\ \bibnamefont {de~Graaf}}, \bibinfo {author} {\bibfnamefont
  {J.~W.~O.}\ \bibnamefont {Garmon}}, \bibinfo {author} {\bibfnamefont {W.~D.}\
  \bibnamefont {Kalfus}}, \bibinfo {author} {\bibfnamefont {Y.}~\bibnamefont
  {Lu}}, \bibinfo {author} {\bibfnamefont {A.}~\bibnamefont {Maiti}}, \bibinfo
  {author} {\bibfnamefont {K.}~\bibnamefont {Sahay}}, \bibinfo {author}
  {\bibfnamefont {N.}~\bibnamefont {Thakur}}, \bibinfo {author} {\bibfnamefont
  {T.}~\bibnamefont {Tsunoda}}, \bibinfo {author} {\bibfnamefont {S.~H.}\
  \bibnamefont {Xue}}, \bibinfo {author} {\bibfnamefont {L.}~\bibnamefont
  {Frunzio}}, \bibinfo {author} {\bibfnamefont {S.~M.}\ \bibnamefont {Girvin}},
  \bibinfo {author} {\bibfnamefont {S.}~\bibnamefont {Puri}},\ and\ \bibinfo
  {author} {\bibfnamefont {R.~J.}\ \bibnamefont {Schoelkopf}},\ }\bibfield
  {title} {\emph {\bibinfo {title} {Dual-rail encoding with superconducting
  cavities}},\ }\href {https://doi.org/10.1073/pnas.2221736120} {\bibfield
  {journal} {\bibinfo  {journal} {Proc. Natl. Acad. Sci.}\ }\textbf {\bibinfo
  {volume} {120}},\ \bibinfo {pages} {41} (\bibinfo {year} {2023})}\BibitemShut
  {NoStop}%
\bibitem [{\citenamefont {Chou}\ \emph {et~al.}(2024)\citenamefont {Chou},
  \citenamefont {Shemma}, \citenamefont {McCarrick}, \citenamefont {Chien},
  \citenamefont {Teoh}, \citenamefont {Winkel}, \citenamefont {Anderson},
  \citenamefont {Chen}, \citenamefont {Curtis}, \citenamefont {de~Graaf} \emph
  {et~al.}}]{Chou2024}%
  \BibitemOpen
  \bibfield  {author} {\bibinfo {author} {\bibfnamefont {K.~S.}\ \bibnamefont
  {Chou}}, \bibinfo {author} {\bibfnamefont {T.}~\bibnamefont {Shemma}},
  \bibinfo {author} {\bibfnamefont {H.}~\bibnamefont {McCarrick}}, \bibinfo
  {author} {\bibfnamefont {T.-C.}\ \bibnamefont {Chien}}, \bibinfo {author}
  {\bibfnamefont {J.~D.}\ \bibnamefont {Teoh}}, \bibinfo {author}
  {\bibfnamefont {P.}~\bibnamefont {Winkel}}, \bibinfo {author} {\bibfnamefont
  {A.}~\bibnamefont {Anderson}}, \bibinfo {author} {\bibfnamefont
  {J.}~\bibnamefont {Chen}}, \bibinfo {author} {\bibfnamefont {J.~C.}\
  \bibnamefont {Curtis}}, \bibinfo {author} {\bibfnamefont {S.~J.}\
  \bibnamefont {de~Graaf}}, \emph {et~al.},\ }\bibfield  {title} {\emph
  {\bibinfo {title} {A superconducting dual-rail cavity qubit with
  erasure-detected logical measurements}},\ }\href
  {https://doi.org/10.1038/s41567-024-02539-4} {\bibfield  {journal} {\bibinfo
  {journal} {Nat. Phys.}\ }\textbf {\bibinfo {volume} {20}},\ \bibinfo {pages}
  {1454} (\bibinfo {year} {2024})}\BibitemShut {NoStop}%
\bibitem [{\citenamefont {Levine}\ \emph {et~al.}(2024)\citenamefont {Levine},
  \citenamefont {Haim}, \citenamefont {Hung}, \citenamefont {Alidoust},
  \citenamefont {Kalaee}, \citenamefont {DeLorenzo}, \citenamefont {Wollack},
  \citenamefont {Arrangoiz-Arriola}, \citenamefont {Khalajhedayati},
  \citenamefont {Sanil} \emph {et~al.}}]{Levine2024}%
  \BibitemOpen
  \bibfield  {author} {\bibinfo {author} {\bibfnamefont {H.}~\bibnamefont
  {Levine}}, \bibinfo {author} {\bibfnamefont {A.}~\bibnamefont {Haim}},
  \bibinfo {author} {\bibfnamefont {J.}~\bibnamefont {Hung}}, \bibinfo {author}
  {\bibfnamefont {N.}~\bibnamefont {Alidoust}}, \bibinfo {author}
  {\bibfnamefont {M.}~\bibnamefont {Kalaee}}, \bibinfo {author} {\bibfnamefont
  {L.}~\bibnamefont {DeLorenzo}}, \bibinfo {author} {\bibfnamefont
  {E.}~\bibnamefont {Wollack}}, \bibinfo {author} {\bibfnamefont
  {P.}~\bibnamefont {Arrangoiz-Arriola}}, \bibinfo {author} {\bibfnamefont
  {A.}~\bibnamefont {Khalajhedayati}}, \bibinfo {author} {\bibfnamefont
  {R.}~\bibnamefont {Sanil}}, \emph {et~al.},\ }\bibfield  {title} {\emph
  {\bibinfo {title} {Demonstrating a long-coherence dual-rail erasure qubit
  using tunable transmons}},\ }\href
  {https://doi.org/10.1103/physrevx.14.011051} {\bibfield  {journal} {\bibinfo
  {journal} {Phys. Rev. X}\ }\textbf {\bibinfo {volume} {14}},\ \bibinfo
  {pages} {011051} (\bibinfo {year} {2024})}\BibitemShut {NoStop}%
\bibitem [{\citenamefont {Koottandavida}\ \emph {et~al.}(2024)\citenamefont
  {Koottandavida}, \citenamefont {Tsioutsios}, \citenamefont {Kargioti},
  \citenamefont {Smith}, \citenamefont {Joshi}, \citenamefont {Dai},
  \citenamefont {Teoh}, \citenamefont {Curtis}, \citenamefont {Frunzio},
  \citenamefont {Schoelkopf},\ and\ \citenamefont
  {Devoret}}]{Koottandavida2024}%
  \BibitemOpen
  \bibfield  {author} {\bibinfo {author} {\bibfnamefont {A.}~\bibnamefont
  {Koottandavida}}, \bibinfo {author} {\bibfnamefont {I.}~\bibnamefont
  {Tsioutsios}}, \bibinfo {author} {\bibfnamefont {A.}~\bibnamefont
  {Kargioti}}, \bibinfo {author} {\bibfnamefont {C.~R.}\ \bibnamefont {Smith}},
  \bibinfo {author} {\bibfnamefont {V.~R.}\ \bibnamefont {Joshi}}, \bibinfo
  {author} {\bibfnamefont {W.}~\bibnamefont {Dai}}, \bibinfo {author}
  {\bibfnamefont {J.~D.}\ \bibnamefont {Teoh}}, \bibinfo {author}
  {\bibfnamefont {J.~C.}\ \bibnamefont {Curtis}}, \bibinfo {author}
  {\bibfnamefont {L.}~\bibnamefont {Frunzio}}, \bibinfo {author} {\bibfnamefont
  {R.~J.}\ \bibnamefont {Schoelkopf}},\ and\ \bibinfo {author} {\bibfnamefont
  {M.~H.}\ \bibnamefont {Devoret}},\ }\bibfield  {title} {\emph {\bibinfo
  {title} {Erasure detection of a dual-rail qubit encoded in a double-post
  superconducting cavity}},\ }\href
  {https://doi.org/10.1103/physrevlett.132.180601} {\bibfield  {journal}
  {\bibinfo  {journal} {Phys. Rev. Lett.}\ }\textbf {\bibinfo {volume} {132}},\
  \bibinfo {pages} {180601} (\bibinfo {year} {2024})}\BibitemShut {NoStop}%
\end{thebibliography}
\end{document}


\title{Supplementary Material for ``Synthetic $\pi$-flux system in 2D superconducting qubit array with tunable coupling"}
\maketitle

\setcounter{equation}{0}
\setcounter{figure}{0}
\setcounter{table}{0}

\newcommand{\figtitle}[1]{{#1}}
\newcommand{\subfiglabel}[1]{(\MakeLowercase{#1})}
\newcommand{\rsubfiglabel}[1]{(\MakeLowercase{#1})}
\newcommand{\subpart}[1]{\textit{#1}}

\newcommand{\rfig}[1]{Fig.~\textcolor{colorref}{\ref{#1}}}
\newcommand{\rfigs}[1]{Figs.~\textcolor{colorref}{\ref{#1}}}

\newcommand{\req}[1]{Eq.~\textcolor{colorref}{(\ref{#1})}}
\newcommand{\rsubfig}[2]{Fig.~\textcolor{colorref}{\ref{#1}}{(\MakeLowercase{#2})}}
\newcommand{\rsubfigs}[3]{Fig.~\textcolor{colorref}{\ref{#1}}{(\MakeLowercase{#2})-(\MakeLowercase{#3})}}
\newcommand{\rsubfigsA}[3]{Fig.~\textcolor{colorref}{\ref{#1}}{(\MakeLowercase{#2}),(\MakeLowercase{#3})}}

\renewcommand{\theequation}{S\arabic{equation}}
\renewcommand{\thefigure}{S\arabic{figure}}
\renewcommand{\thetable}{S\arabic{table}}

\renewcommand{\figurename}{Fig.}
\def\figureautorefname{Fig.}

\tableofcontents
\newpage

\section{Details on theoretical model}

The system discussed in the main text can be described by an effective Hamiltonian of a rhombic lattice model with $L=3l+1$ sites ($l$ is the number of rhombic plaquettes)
\cite{Longhi2021,Li2022,Martinez2023,Rosen2024}:
\begin{equation}
H/\hbar = \sum_{i,j} \Delta_{i,j} \sigma_{i,j}^{+} \sigma_{i,j}^{-}   - J \sum_{j=1}^{l} \left[
\sigma_{A,j}^{+} \left( \sigma_{\uparrow,j}^{-} +  \sigma_{\downarrow,j}^{-} \right) 
+ \sigma_{A,j+1}^{+} \left(  \sigma_{\uparrow,j}^{-} + e^{i\Phi}  \sigma_{\downarrow,j}^{-} \right)
+ \mathrm{H.c.}\right],
\end{equation}
where $\sigma^{+}_{i,j}$ ($\sigma_{i,j}^{-}$) represents the raising (lowering) operator for the qubit at site $(i,j)$ with $i\in \left\{A,\uparrow,\downarrow\right\}$,
$\sigma_{i,j}^{+} \vert 0\rangle^{\otimes L} = \vert 1_{i,j}\rangle $,
$\Delta_{i,j}/2\pi$ is the detuning of the qubit relative to the average system frequency,
and $J$ is the homogeneous amplitude of NN coupling strengths between qubits.

When the system is configured as $\Phi=0$ by using the same negative coupling strengths and $\Delta_{i,j}=0$,
the effective Hamiltonian in the single-excitation subspace can be written in the form of one-dimensional (1D) array:
\begin{equation}
H_{\Phi=0}/\hbar = -\sqrt{2}J \sum_{j=1}^{l} \left(\sigma_{A,j}^{+} \sigma_{+,j}^{-}+\sigma_{A,j+1}^{+} \sigma_{+,j}^{-}+\mathrm{H.c.} \right),
\end{equation}
where $\sigma_{+,j}^{+} = \left(\sigma_{\uparrow,j}^{+} + \sigma_{\downarrow,j}^{+} \right)/\sqrt{2}$,
$\sigma_{-,j}^{+} = \left(\sigma_{\uparrow,j}^{+} -\sigma_{\downarrow,j}^{+} \right)/\sqrt{2}$.
In this representation, the Bell state $\vert +_{j} \rangle =$ $(\vert 1_{\uparrow,j} \rangle + $ 
$\vert 1_{\downarrow,j} )/\sqrt{2}$ forms an effective 1D array on the $2l+1$ sites $\{ (A,1),(+,1)$ $,(A,2),(+,2),\cdots \}$ in the single-excitation subspace~\cite{Danieli2021}.
Besides the effective 1D array featuring two dispersive energy bands,
the Bell state $\vert -_{j} \rangle = \left(\lvert 1_{\uparrow,j} \rangle - \lvert 1_{\downarrow,j} \rangle \right)/\sqrt{2}$ is decoupled with the system Hamiltonian, 
$H \vert -_{j} \rangle = 0$, resulting in a flat band with zero eigenenergies.

In another scenario of $\Phi=\pi$, the effective Hamiltonian in the single-excitation subspace can be written as:
\begin{equation}
H_{\Phi=\pi}/\hbar = -\sqrt{2}J \sum_{j=1}^{l} \left(
\sigma_{A,j}^{+} \sigma_{+,j}^{-}+\sigma_{A,j+1}^{+} \sigma_{-,j}^{-}+\mathrm{H.c.}
\right).
\end{equation}
This yields decoupled three-level systems in the bulk sites $\{ (-,j-1),(A,j),(+,j)\}$,
and two-level systems in the edge sites $\{(A,0),(+,0)\}$ and $\{(-,l),(A,l+1)\}$.

Introducing the anti-symmetric detunings ($\Delta_{\uparrow,j}=\Delta$ and $\Delta_{\downarrow,j}=-\Delta$) on the qubit,
the system can be described by the Hamiltonian
\begin{equation}
H/\hbar = \sum_{j} \Delta  (
\sigma^{+}_{\uparrow,j}\sigma^{-}_{\uparrow,j} -\sigma^{+}_{\downarrow,j}\sigma^{-}_{\downarrow,j})  
- J \sum_{j=1}^{l} \left[
\sigma_{A,j}^{+} \left( \sigma_{\uparrow,j}^{-} +  \sigma_{\downarrow,j}^{-} \right) 
+ \sigma_{A,j+1}^{+} \left(  \sigma_{\uparrow,j}^{-} \pm  \sigma_{\downarrow,j}^{-} \right)
+ \mathrm{H.c.}\right].
\end{equation}
In the basis $\vert \pm_{j} \rangle = \left(\vert 1_{\uparrow,j} \rangle \pm \vert 1_{\downarrow,j} \right)/\sqrt{2}$,
the Hamiltonian can be written as
\begin{equation}
H/\hbar = \sum_{j=1}^l \left[
\Delta \sigma^{+}_{+,j}\sigma^{-}_{-,j} 
- \sqrt{2} J \left(
\sigma_{A,j}^{+} \sigma_{+,j}^{-}+\sigma_{A,j+1}^{+} \sigma_{\pm,j}^{-}\right)
+ \mathrm{H.c.}\right],
\end{equation}
which gives additional coupling between the states $\vert 1_{+,j} \rangle$ and $\vert 1_{-,j} \rangle$.
For the $\Phi=\pi$ rhombic system, the anti-symmetric detunings induce inter-cell couplings between the effective unit cell of three-level systems, comprising the states
$\{ \vert 1_{-,j-1}\rangle, \vert 1_{A,j} \rangle, \vert 1_{+,j} \rangle \}$ localized in the bulk sites,
and two-level systems with states $\{ \vert 1_{A,0}\rangle, \vert 1_{+,0} \rangle\}$ or $\{ \vert 1_{-,l}\rangle, \vert 1_{A,l+1} \rangle\}$ localized at the edge sites.
Consequently, the $\pi$-flux system with anti-symmetric detunings can be effectively mapped onto a 1D trimer lattice \cite{MartinezAlvarez2019,Anastasiadis2022}, characterized by intra-cell couplings of $\sqrt{2}J$ and inter-cell couplings of $\Delta$.

\section{Experiment}
\subsection{Device information}
Our experiment is performed on a $6 \times 11$ square-lattice quantum processor using tunable coupling scheme~\cite{Yan2018,Xu2020},
which shares the same overall design as described in Ref.~\cite{liang2024dephasing}.
The effective coupling strength between two neighboring qubits is tunable within a range of about $+5$ MHz to $-20$ MHz.
In this study, we utilize 7 $($or 4$)$ qubits to construct rhombic plaquettes for investigating quantum properties.
These qubits are labeled as $Q_{i, j}$ $(i=A, \uparrow, \downarrow \text{ and } j=1, 2)$.
The detailed characterization parameters of these qubits are provided in \rtab{Tab01}.

\begin{table}[!t]
    \centering
    \caption{
    \label{Tab01}
    \figtitle{Qubits properties.}
    $\omega_{\textrm{min}}/2\pi$ ($\omega_{\textrm{max}}/2\pi$) is the qubit sweet point and $\omega_{\textrm{idle}}/2\pi$ is the qubit idle frequency. $T_{1, \textrm{idle}}$ is the qubit relaxation time measured at idle frequency. $T_{2, \phi}$ denotes the qubit dephasing time. $F_{0}$ and $F_{1}$ are the readout fidelities of qubit $|0\rangle$ state and $|1\rangle$ state, respectively.}
    \begin{tabular}{p{5.0cm}<{\centering}p{1.5cm}<{\centering}p{1.5cm}<{\centering}p{1.5cm}<{\centering}p{1.5cm}<{\centering}p{1.5cm}<{\centering}p{1.5cm}<{\centering}p{1.5cm}<{\centering}p{1.5cm}<{\centering}p{1.5cm}<{\centering}p{1.5cm}<{\centering}p{1.5cm}<{\centering}}
     \toprule
      & $Q_{A, 1}$ & $Q_{\uparrow, 1}$ & $Q_{\downarrow, 1}$ & $Q_{A, 2}$ & $Q_{\uparrow, 2}$ & $Q_{\downarrow, 2}$ &$Q_{A, 3}$\\
     \midrule
     $\omega_{\textrm{min}}/2\pi$ (GHz) & 3.639 & 3.816 & 3.972 & 3.795 & 3.909 & 3.952 & 3.898\\
     $\omega_{\textrm{max}}/2\pi$ (GHz) & 4.891 & 4.867 & 5.055 & 5.121 & 5.109 & 5.217 & 5.083\\
     $\omega_{\textrm{idle}}/2\pi$ (GHz) & 4.120 & 4.160 & 4.239 & 4.208 & 4.110 & 4.165 & 4.231 \\
     $\omega_{\textrm{r}}/2\pi$ (GHz) & 6.216 & 6.112 & 6.118 & 6.160 & 6.186 & 6.190 & 6.083 \\
     $T_{1, \textrm{idle}}$ ($\mu$s) & 63.7 & 68.4 & 54.0 & 40.0 & 55.6 & 42.8 & 68.8 \\
     $T_{2, \phi}$ ($\mu$s) & 9.7 & 12.8 & 9.8 & 3.8 & 12.8 & 8.4 & 10.8 \\
     $F_{0}$ & 0.983 & 0.976 & 0.972 & 0.990 & 0.987 & 0.989 & 0.975 \\
     $F_{1}$ & 0.973 & 0.972 & 0.965 & 0.962 & 0.940 & 0.972 & 0.972 \\
     \bottomrule
\end{tabular} 
\end{table}

\begin{reviseEnv}

\subsection{Details of tunable couplers}

In this section, we provide detailed information on tunable couplers that enable the $\pi$-flux configuration.
For the coupler designs in the experiments, we use the floating transmon~\cite{Sete2021} with two Josephson junctions in our device.
The qubits and couplers in the device have two Josephson junctions, which enable the flux control mechanism of their frequencies $\omega(\Phi_{\mathrm{ext}})$ by applying an external flux $\Phi_{\mathrm{ext}}$ through the control line~\cite{Blais2021}.
Each floating transmon qubit is surrounded by four tunable couplers, and the two transmon qubits are capacitively coupled to a floating tunable coupler~\cite{Yang2024}.

The effective coupling strength between qubits $Q_A$ and $Q_B$ can be controlled by applying external flux on the corresponding coupler $C$ coupled to $Q_A$ and $Q_B$~\cite{Yan2018,Xu2020}.
The system of qubits $Q_{A,B}$ and coupler $C$ can be described by the Hamiltonian
\begin{equation}
\begin{aligned}
	H/\hbar= & \sum_{i=A,B,C}\left(\omega_i a_i^{\dagger} a_i+\frac{U_i}{2} a_i^{\dagger} a_i^{\dagger} a_i a_i\right) +
	g_{AC}\left(a_A^{\dagger}a_C+a_A a_C^{\dagger}\right)+
	g_{BC}\left(a_B^{\dagger}a_C+a_B a_C^{\dagger}\right) \\
    &+
    g_{AB}\left(a_A^{\dagger}a_B+a_A a_B^{\dagger}\right), 
\end{aligned}
\end{equation}
where $a_i$ ($a^\dagger_i$) is the annihilation (creation) operator, $\omega_i$ is the qubit or coupler frequency, 
$U_i$ is the qubit or coupler anharmonicity, 
$g_{AC}$ ($g_{BC}$) is the coupling strength between $Q_{A}$ ($Q_{B}$) and coupler $C$, 
and $g_{AB}$ is the coupling strength between $Q_{A}$ and $Q_{B}$.
When the coupler frequency is largely detuned from the qubits frequencies $\vert \omega_{A,B} - \omega_C \vert \gg g_{AC},g_{BC}$, the effective Hamiltonian of the system is given as
\begin{equation}
H_{\mathrm{eff}}/\hbar=  \sum_{i=A,B}\left(\omega_i a_i^{\dagger} a_i+\frac{U_i}{2} a_i^{\dagger} a_i^{\dagger} a_i a_i\right) + g_{\mathrm{eff}}\left(a_A^{\dagger}a_B+a_A a_B^{\dagger}\right),
\end{equation}
where the effective coupling strength~\cite{Yan2018}
\begin{equation}
g_{\mathrm{eff}} \approx 
g_{AB}+\frac{ g_{AC} g_{BC}}{2}\left(
\frac{1}{\omega_A-\omega_C}+\frac{1}{\omega_B-\omega_C}
\right),
\end{equation}
can be modulated by tuning the coupler frequency $\omega_C$ through the flux control (Z control) line of the coupler~\cite{Yan2018,Xu2020}.
In the experiment, we can apply a Z control pulse on the coupler with different Z pulse amplitudes (zpa) to change the effective coupling strength $g_\mathrm{eff}$, 
which can be tuned into either positive $g_\mathrm{eff}>0$ or negative $g_\mathrm{eff}<0$.
For example, the coupling strength $g_\mathrm{eff}\approx g_{AB} >0$ when the coupler is largely detuned from the qubit frequencies $\omega_C \gg \omega_{A,B}$.
The coupling-off point $g_\mathrm{eff}=0$ can also be used to identify the boundary between the positive and negative coupling strength~\cite{Yan2018}.
When the four qubits are coupled in a sequential manner of the closed loop, which forms a plaquette with four sites, the system can be described by a generic Hamiltonian with $N=4$
\begin{equation}
H = \sum_{n=1}^{N-1} J_n (a^\dagger_{n} a_{n+1} + a_{n} a^\dagger_{n+1}) + J_N (a^\dagger_{N} a_{1} + a_{1} a^\dagger_{N}),
\end{equation}
where $J_n = \vert J_n\vert e^{i\varphi_n} $ is the complex-valued coupling strength. 
Here we can assign $e^{i\varphi_n} = +1$ ($e^{i\varphi_n} = -1$) by tuning the coupling strength as positive (negative), and the synthetic flux $\Phi \equiv \sum_n \varphi_n$.
For the zero-flux configuration, we can modulate all coupling strengths to be negative by controlling the frequencies of the tunable couplers.
For the $\pi$-flux configuration, we can modulate one of the coupling strengths to be positive, and three of the other coupling strengths to be negative.

\subsection{Crosstalk in qubits array}

When controlling the frequencies in the qubits array $Q_i$, the crosstalk effect will change the applied flux $V_i$ into $V_i'$ when the flux pulse reaches the targeted qubit $Q_i$.
The Z control line crosstalk can be estimated by measuring the linear response with respect to the flux amplitudes,
where the crosstalk matrix $M$ gives $V_i' = M V_i$ and can be used to correct the crosstalk by multiplying the inverse of the matrix by the values of flux amplitudes.
The crosstalk between the source and target qubit can be extracted by the linear response of the Z pulse amplitudes (zpa),
where a driving pulse with fixed frequency $\omega_d$ is applied through the XY control line of the target qubit, and the Z pulses of the target and source qubits are simultaneously applied.
The qubit is excited if the flux pulses tune the qubit frequency into $\omega_d$,
which leads to an increased measured population when scanning the values of source and target qubit zpa,
as shown in \rfig{FigS_crosstalk}.
By fitting the linear response with respect to the flux amplitudes, we extract the strength of the crosstalk element as $\sim 0.06\%$.
The strength of the crosstalk effect is largely related to the fabricated device. 
The superconducting quantum processor used in this work comprises a top chip and a bottom carrier chip using the flip-chip technique~\cite{Satzinger2018}.
To mitigate the signal crosstalk from the aspect of chip fabrication, we create SiO$_2$-supported bridges on the bottom chip to shield critical circuits especially the control lines.

\begin{figure*}[!t]
	\centering
    \includegraphics[width=0.65\textwidth]{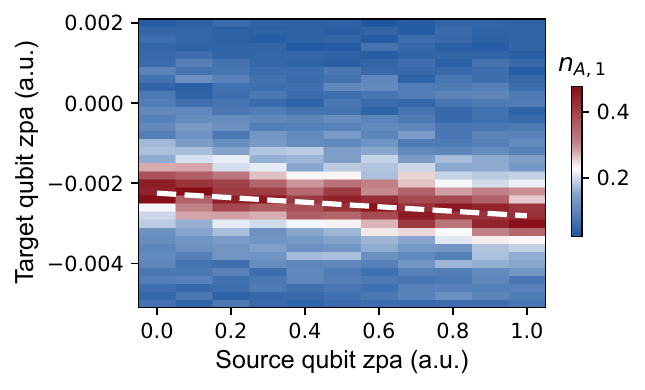}
    \revise{
	\caption{\label{FigS_crosstalk}
        \figtitle{Measurement of Z crosstalk between the source and target qubit.}
        The dashed line is the result of linear fitting to obtain the crosstalk matrix element.
	}
    }
\end{figure*}

\subsection{Experimental benchmarks of $\pi$-flux condition}

\begin{figure*}[!t]
    \centering
    \includegraphics[width=0.5\textwidth]{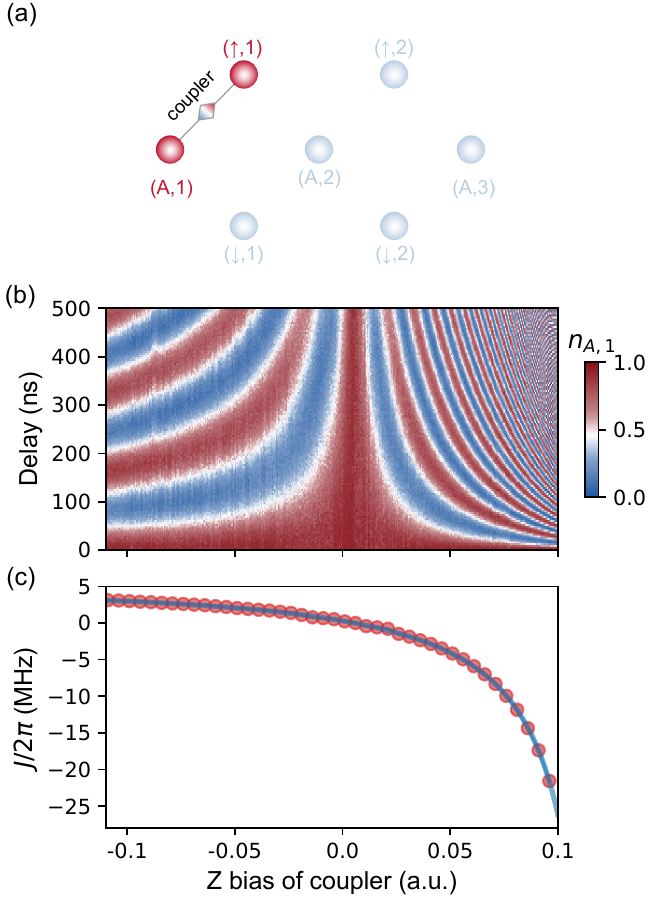}
    \revise{
	\caption{\label{FigS_swap2d}
        \figtitle{Characterization of the tunable coupling.}
        \subfiglabel{a}
        Schematic of the two-qubit system for characterizing the tunable coupling, where the other shaded qubits (marked as blue) are largely detuned.
        \subfiglabel{b}
        Vacuum Rabi oscillation between the two qubits at different coupler bias, where the colors denote the population $n_{A,1}$.
        \subfiglabel{c}
        The tunable coupling strength extracted from \subfiglabel{b}.
	}
    }
\end{figure*}

As shown in \rsubfig{FigS_swap2d}{a}, we firstly perform the characterization of the tunable coupling between each coupler to calibrate the parameters for the desired positive or negative coupling strength used in the following measurements.
We perform vacuum Rabi oscillation between the first excited states of the two qubits at different coupler bias to characterize the effective coupling strength as shown in \rsubfigs{FigS_swap2d}{b}{c}.

To benchmark the $\pi$-flux condition in the qubits array, we start from the calibration of the minimal rhombic array with only one plaquette.
As shown in \rsubfig{FigS_squareEvo}{a}, we choose the four-qubit subsystem by detuning the other qubits and configure the positive or negative coupling strengths $\pm J$ by tunable couplers.
The single-excitation subspace of the four qubits comprising the states $\{\vert 1_{A,1}\rangle,\vert 1_{\uparrow,1}\rangle,\vert 1_{A,2}\rangle,\vert 1_{\downarrow,1}\rangle\}$ can be described by the Hamiltonian in a matrix form
\begin{equation}
H = -J
    \begin{pmatrix}
    0 & 1 & 0 & 1 \\
    1 & 0 & 1 & 0 \\
    0 & 1 & 0 & e^{-i\Phi} \\
    1 & 0 & e^{i\Phi} & 0 \\
    \end{pmatrix}.
\end{equation}
Here we focus on the cases of zero-flux $\Phi=0$ and $\pi$-flux $\Phi=\pi$.

\begin{figure*}[!b]
    \centering
    \includegraphics[width=0.65\textwidth]{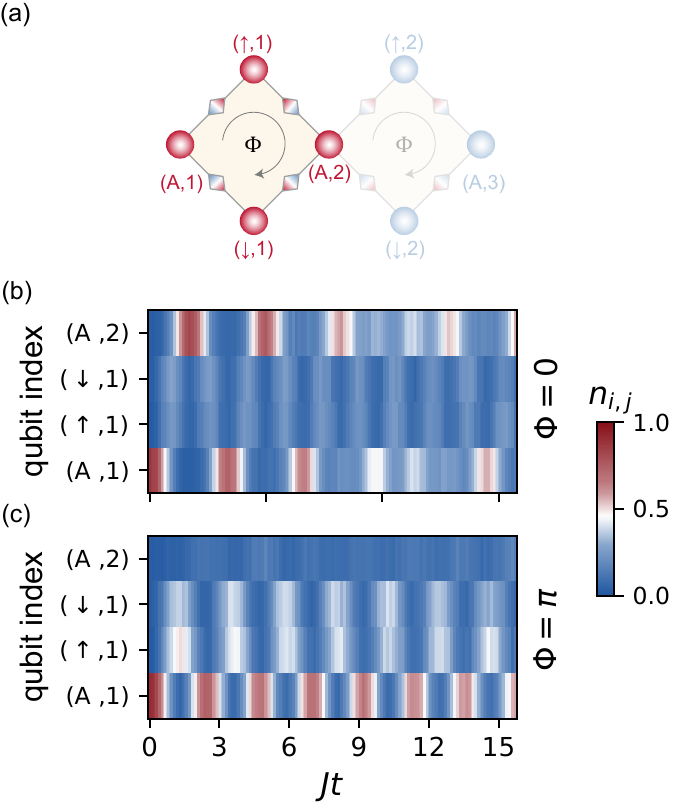}
    \revise{
	\caption{\label{FigS_squareEvo}
        \figtitle{Characteristic dynamics of the four-qubit system.}
        \subfiglabel{a}
        Schematic of the four-qubit system marked as red in the left plaquette, where the other shaded qubits (marked as blue) are largely detuned.
        \subfiglabel{b}-\subfiglabel{c}
        Measured population dynamics $n_{i,j}$ versus evolution time $Jt$ for \subfiglabel{b} $\Phi=0$ and \subfiglabel{c} $\Phi=\pi$ with the initial state $\vert 1_{A,1}\rangle$.
	}
    }
\end{figure*}

For the zero-flux $\Phi=0$ condition, the system evolves as $\vert \psi (t)\rangle = e^{-i Ht} \vert \psi_0 \rangle$ with the initial state $\vert \psi_0 \rangle$, where the unitary operator $U\equiv e^{-i Ht}$ is given as 
\begin{equation}
U=
\left(
\begin{array}{cccc}
 \cos ^2 (J t) & -i \sin(J t) \cos(J t) & -\sin^2 (J t) & -i\sin(J t) \cos(J t) \\
 -i\sin(J t) \cos(J t) & \cos ^2 (J t) & -i\sin(J t) \cos(J t) & -\sin ^2 (J t) \\
 -\sin^2 (J t) & -i\sin(J t) \cos(J t) & \cos ^2 (J t) & -i\sin(J t) \cos(J t) \\
 -i\sin(J t) \cos(J t) & -\sin ^2 (J t) & -i\sin(J t) \cos(J t) & \cos^2 (J t) \\
\end{array}
\right)
.
\end{equation}
The population dynamics of evolved state $\vert \psi(t)\rangle$ from a given initial state $\vert \psi_0 \rangle = \vert 1_{A,1}\rangle$ is 
\begin{equation}
\begin{split}
 \{n_{A,1} ,n_{\uparrow,1}, n_{A,2} , n_{\downarrow,1} \}= 
 \{
 \cos^4(Jt),\sin^2(2Jt)/4 , \sin^4(Jt),\sin^2(2Jt)/4
 \},
\end{split}
\end{equation}
showing an oscillated population dynamics $n_{A,2} = \sin^4(Jt)$ and $\vert \psi (t)\rangle $ evolved into $\vert 1_{A,2}\rangle$ when $Jt=\pi/2$.
For the $\pi$-flux $\Phi=\pi$ condition, the system evolves as $\vert \psi (t)\rangle = e^{-i Ht} \vert \psi_0 \rangle$ with the unitary operator $U\equiv e^{-i Ht}$ given as
\begin{equation}
U= 
\frac{1}{\sqrt{2}}
\left(
\begin{array}{cccc}
 \sqrt{2}\cos  \left(\sqrt{2} J t\right) & -i \sin  \left(\sqrt{2} J t\right) & 0 & -i \sin  \left(\sqrt{2} J t\right) \\
 -i \sin  \left(\sqrt{2} J t\right) & \sqrt{2} \cos  \left(\sqrt{2} J t\right) & -i \sin  \left(\sqrt{2} J t\right) & 0 \\
 0 & -i  \sin  \left(\sqrt{2} J t\right) & \sqrt{2} \cos  \left(\sqrt{2} J t\right) & i \sin  \left(\sqrt{2} J t\right) \\
 -i \sin  \left(\sqrt{2} J t\right) & 0 & i \sin  \left(\sqrt{2} J t\right) & \sqrt{2} \cos  \left(\sqrt{2} J t\right) \\
\end{array}
\right)
.
\end{equation}
The population dynamics of evolved state $\vert \psi(t)\rangle$ from a given initial state $\vert \psi_0 \rangle = \vert 1_{A,1}\rangle$ is 
\begin{equation}
\begin{split}
 \{n_{A,1} ,n_{\uparrow,1}, n_{A,2} , n_{\downarrow,1} \}= 
 \{
 \cos^2(\sqrt{2}Jt),\sin^2(\sqrt{2}Jt)/2 , 0,\sin^2(\sqrt{2}Jt)/2
 \},
\end{split}
\end{equation}
showing an vanishing population dynamics $n_{A,2} =0$ due to destructive interference of $\pi$-flux.

Therefore, we can experimentally benchmark the characteristic dynamics for the zero-flux and $\pi$-flux conditions generated by the configuration of positive and negative coupling strength.
Since the coupling strength discussed here only involves positive or negative values due to the tunable coupling scheme, the case of other complex numbers of $\Phi$ besides $\Phi=0,\pi$ can be safely avoided in the experiment.
We choose the four-qubit subsystem by detuning the other qubits and configure the positive or negative coupling strengths $\pm J$ by tunable couplers as shown in \rsubfig{FigS_squareEvo}{a}.
By initially preparing the qubit at site $(A,1)$ in its excited state $\vert 1_{A,1}\rangle$, we examine the characteristic population dynamics of the four-qubit subsystem for both zero-flux and $\pi$-flux conditions.
As shown \rsubfigs{FigS_squareEvo}{b}{c}, the measured results give an oscillated population dynamics of $n_{A,2}$ and maximum value  at $Jt\approx \pi/2$ for $\Phi=0$,
while the population hardly evolve into the site $(A,2)$ for $\Phi=\pi$ condition.
By performing the calibration and benchmarks procedures described above for each plaquette in the rhombic array, we can generate the larger rhombic array of superconducting qubits for both zero-flux $\Phi=0$ and $\pi$-flux $\Phi=\pi$.

\end{reviseEnv}

\subsection{Experimental realization and numerical simulation}

\begin{figure*}[!h]
	\centering
    \includegraphics[width=0.95\textwidth]{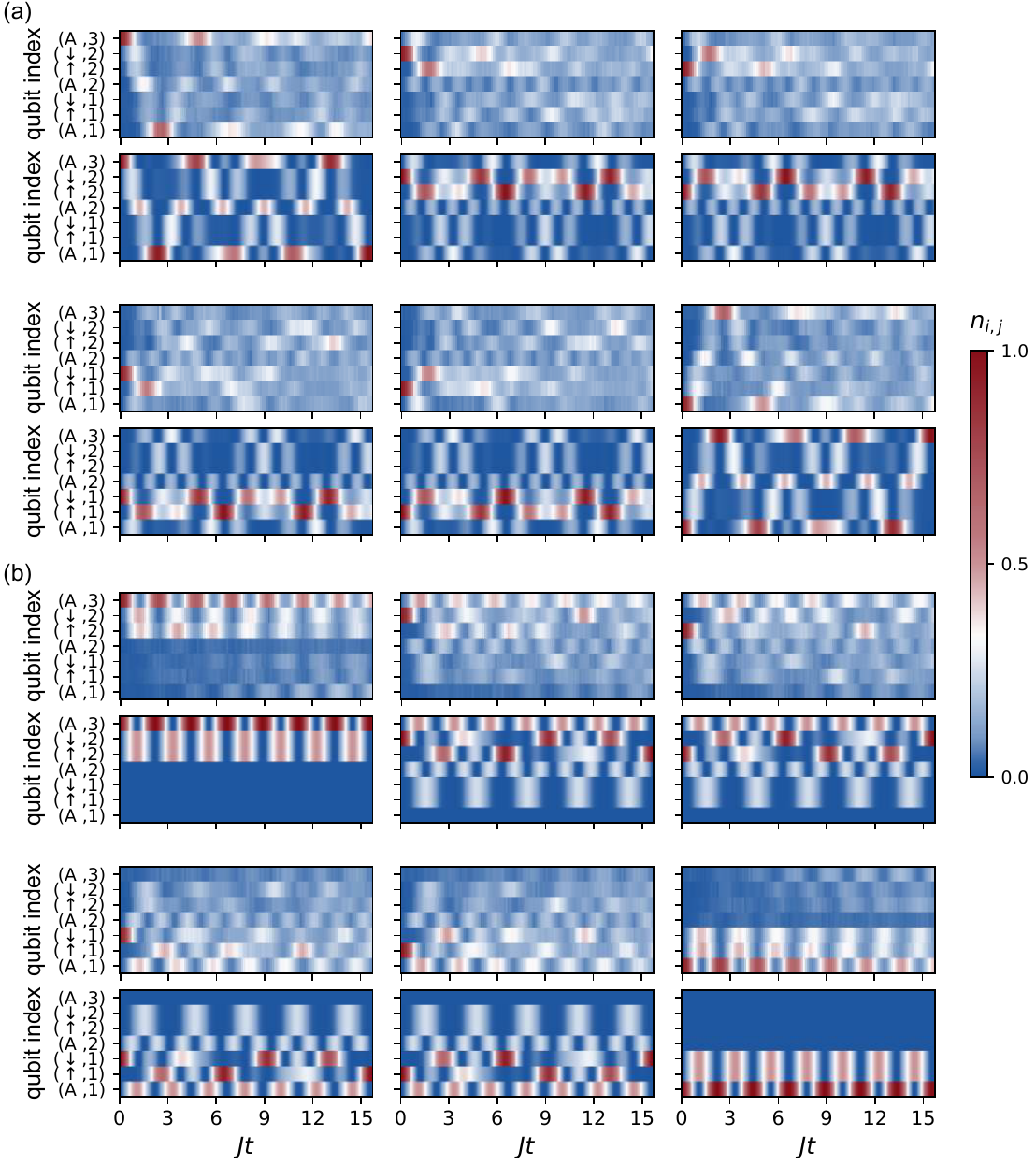}
	\caption{\label{figS1}
        \figtitle{Characteristic dynamics of the rhombic array with coupler-assisted flux $\Phi=0,\pi$}.
        \subfiglabel{a}-\subfiglabel{b}
        Measured population dynamics $n_j$ versus normalized evolution time $Jt$ for \rsubfiglabel{a} $\Phi=0$ and \rsubfiglabel{b} $\Phi=\pi$. The odd rows show the experimental results, while the even rows present the corresponding theoretical simulation results.
	}
\end{figure*}

\revise{Subsequently}, we investigate the characteristic population dynamics for $L=7$ qubits under sequential excitation.
Specifically, we present the results for excitation at the sites $(A,1)$, $(\uparrow,1)$, $(\downarrow,1)$, $(\uparrow,2)$, $(\downarrow,2)$ and $(A,3)$, respectively, in addition to the previously shown excitation at the site $(A,2)$ in Fig. 2 of the main text.
The experimental results and corresponding numerical simulations are presented in \rfig{figS1} with effective fluxes $\Phi = 0$ and $\Phi=\pi$, using the same coupling strength and the normalized evolution time.

\begin{figure*}[htbp]
	\centering
    \includegraphics[width=0.99\textwidth]{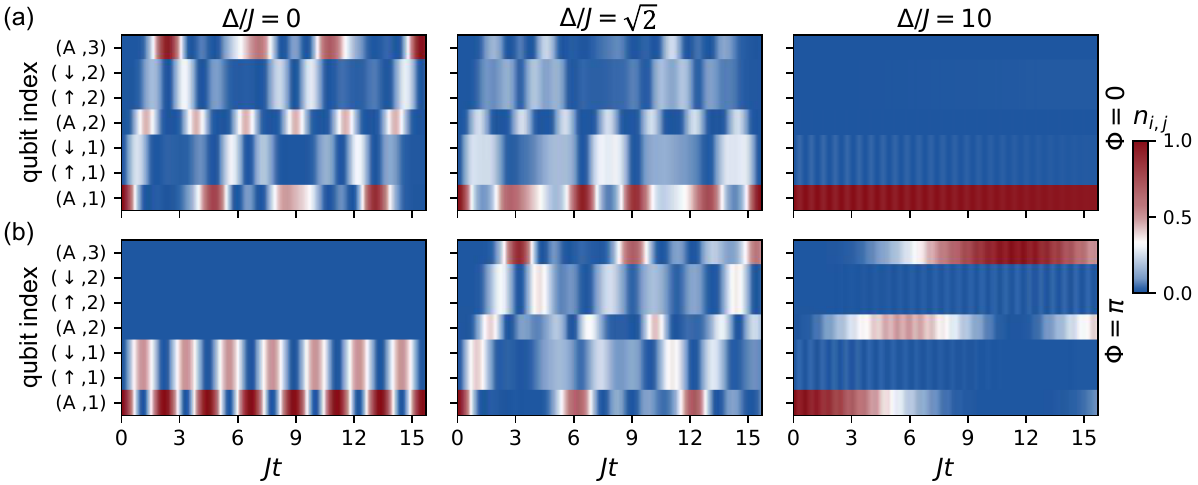}
	\caption{\label{figS2}
		Numerical simulation results of population dynamics $n_j$ for \rsubfiglabel{a} $\Phi=0$ and \rsubfiglabel{b} $\Phi=\pi$,
		with anti-symmetric detuning $\Delta=0$, $\sqrt{2} J$ and $10 J$. The difference between the experimental results and the numerical simulations for the case with $\Delta = 10 J$ arises from the residual next-nearest-neighbor (NNN) couplings.
	}
\end{figure*}

The (de)localized dynamics induced by adding anti-symmetric detunings, $\Delta$ and $-\Delta$, to the qubits at the sites $(\uparrow,j)$ and $(\downarrow,j)$, respectively, is realized through additional z pulses that adjust the frequencies of the target qubits.
Numerical simulations corresponding to the experimental results in Fig. 3 of the main text, with three different detuning strengths, $\Delta = 0, \sqrt{2}J$ and $10J$, are presented in \rfig{figS2} for both the $\Phi = 0$ and $\Phi = \pi$ cases.

\begin{reviseEnv}
The rhombic system in the ground state preparation involves the slow adiabatic evolution,
in which the qubits decoherence during the slow evolution is the dominated error mechanism resulting in the decrease of fidelity compared with the ideal case.
We consider the fidelity $F = \sum_{i,j} \sqrt{n_{i,j}^{\mathrm{calc}} n_{i,j}^{\mathrm{ideal}}}$ for the calculated population distribution $n_{i,j}^{\mathrm{calc}}$ and ideal population distribution $n^{\mathrm{ideal}}_{i,j}$ (without considering decoherence) following Refs.\cite{Yan2019,Shi2023}.
The system evolution is governed by the Lindblad master equation
\begin{equation}\label{eq:lindblad}
	\frac{d\rho}{dt} = -\frac{i}{\hbar} \left[H, \rho\right] + 
	\sum_{j} \left(\mathcal{L}_j \rho \mathcal{L}_j^\dagger - \frac{1}{2} \left\{\mathcal{L}_j^\dagger \mathcal{L}_j ,\rho \right\} \right),
\end{equation}
where $\mathcal{L}_j=\sqrt{\Gamma_j}\sigma_{j}^{+}\sigma_{j}^{-}$ denotes the dephasing collapse operator, $\rho$ is the density matrix, $\Gamma_j$ is the dephasing rate.
We calculate the system evolution using the Lindblad master equation with QuTiP python library~\cite{Johansson2012}.
As shown in \rfig{FigS_fidelity}, we numerically consider the state preparation of the $\pi$-flux rhombic system discussed in Fig. 4 of the main text.
We evaluate the calculated population distribution $n_{i,j}^{\mathrm{calc}}$ considering the dephasing operator $\mathcal{L}_j=\sqrt{\Gamma_j}\sigma_{j}^{+}\sigma_{j}^{-}$ in the Lindblad master equation with $1/\Gamma_i = 1~\mu \mathrm{s}$ and $10~\mu\mathrm{s}$,
which shows a lower fidelity of population distribution when considering a larger dephasing rate (shorter dephasing time).

\begin{figure*}[!b]
    \centering
    \includegraphics[width=0.6\textwidth]{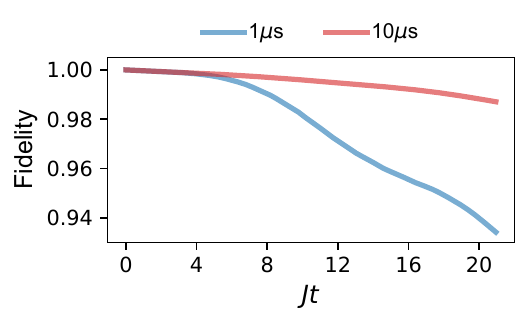}
    \revise{
	\caption{\label{FigS_fidelity}
        Numerical simulation of the fidelity
        for the calculated population distribution $n_{i,j}^{\mathrm{calc}}$ and ideal population distribution $n_{i,j}^{\mathrm{ideal}}$ corresponding to the state preparation discussed in the main text.
        The calculated population $n_{i,j}^{\mathrm{calc}}$ is considered for the $\pi$-flux rhombic system in Fig. 4 of the main text, 
        the dephasing rate $1/\Gamma_i = 1~\mu \mathrm{s}$ and $10~\mu\mathrm{s}$ in the Lindblad master equation of \req{eq:lindblad},
        and $n_{i,j}^{\mathrm{ideal}}$ corresponds to the case without considering the decoherence effects.
	}
    }
\end{figure*}

\end{reviseEnv}

\clearpage